\documentclass[12pt]{article}
\usepackage{amssymb,amsfonts,amsmath,graphicx,longtable}

\newcommand{\be}{\begin{eqnarray*}}
\newcommand{\ee}{\end{eqnarray*}}
\newcommand{\ben}{\begin{eqnarray}}
\newcommand{\een}{\end{eqnarray}}
\newtheorem{prop}{Proposition}
\newcommand{\bprop}{\begin{prop}}
\newcommand{\eprop}{\end{prop}}
\newtheorem{proof}{Proof}
\newcommand{\bproof}{\begin{proof}}
\newcommand{\eproof}{\end{proof}}
\newtheorem{thm}{Theorem}
\newcommand{\bthm}{\begin{thm}}
\newcommand{\ethm}{\end{thm}}

\def\reals{\mathbf{R}}
\def\ints{\mathbf{Z}}

\def\lap{{\mathfrak{n}}}
\def\csa{{\mathfrak{h}}}
\def\lag{{\mathfrak{g}}}

\def\las{{\mathfrak{s}}}

\begin{document}

\begin{titlepage}
\begin{flushright}
hep-th/0309198\\
DAMTP 2003-88\\
RI-09-03\\
KCL-MTH-03-14\\[1cm]
\end{flushright}
\begin{center}
{\bf \Large Very-extended Kac-Moody
algebras and their interpretation at low levels}\\[2cm]
{\large Axel Kleinschmidt\\}
Department of Applied Mathematics and Theoretical Physics,
University of Cambridge, Wilberforce Road, Cambridge CB3 0WA, UK\\
Email: {\tt a.kleinschmidt@damtp.cam.ac.uk}\\[5mm]
{\large Igor Schnakenburg\\}
Racah Institute of Physics, The Hebrew University, Jerusalem 91904,
Israel\\
Email: {\tt igorsc@phys.huji.ac.il}\\[5mm]
{\large Peter West}\\
Department of Mathematics, King's College, London WC2R 2LS, UK\\
Email: {\tt pwest@mth.kcl.ac.uk}\\[5mm]

\mbox{}\\[2cm]

\end{center}

\renewcommand{\abstract}{\begin{center}\bf
Abstract\\[.3cm]\end{center}}

\begin{abstract}
We analyse the very-extended Kac-Moody algebras as
representations in terms of certain $A_d$ subalgebras and find the
generators at low levels.
Our results for low levels
agree precisely with the bosonic field content of the theories
containing gravity, forms and scalars which upon reduction to three
dimensions can be described by a non-linear realisation. We explain
how the Dynkin diagrams of the very-extended algebras encode
information about the field content and generalised  
T-duality transformations. 

\end{abstract}

\end{titlepage}
%
%

\begin{section}{Introduction}

The finite-dimensional Lie algebras  have
played an essential r\^ole in our understanding of particle physics
culminating in the development of the standard model which describes
three of the four forces of nature. In fact, almost all of the
finite-dimensional semi-simple Lie algebras classified by Cartan have
occurred in discussions of either unified theories or space-time
symmetries.  The discovery of Kac-Moody algebras in the 1960s
considerably enlarged the class of Lie algebras. A given Kac-Moody
algebra of rank $r$ is completely specified by a finite-dimensional
integer valued $r\times r$  matrix $A_{ij}$, called the Cartan matrix.
Subject to a small number of properties one can define a Kac-Moody
algebra for any such matrix $A_{ij}$. The class of Kac-Moody algebras
comprises the finite-dimensional semi-simple Lie algebras and the
affine algebras but also more general algebras.

Affine Lie algebras
have played an important role in string theory and conformal field theory
(for a review see 
\cite{GoOl86}. 
Apart from exceptions,  until 
recently no significant r\^ole has been found for more general Kac-Moody
algebras. The two exceptions concerned the dimensional reduction of
supergravity theories and hyperbolic Kac-Moody algebras; it was suggested
\cite{Ju82}
that eleven-dimensional supergravity when dimensionally reduced
to one dimension might possess an $E_{10}$ symmetry and it was shown that
four dimensional $N=1$ supergravity possessed a hyperbolic Kac-Moody
algebra \cite{Ni92}. Relatively recently it  has been found
that the dynamics of theories of the type {\rm II} supergravities 
near a space-like singularity becomes the free motion of a massless
particle with scattering off the walls of the fundamental Weyl chamber
of $E_{10}$, called  cosmological billiards
\cite{cosmbill}. 
  
In the last two years it has become clear that the maximal
supergravity theories in ten and eleven dimensions have Kac-Moody
algebras associated with them. It has been shown that the
bosonic sector of eleven-dimensional supergravity can be expressed as a
non-linear realisation \cite{We00}. Although the  algebra used in this
non-linear realisation was not a Kac-Moody algebra it was argued
that there should exist an extension of eleven-dimensional supergravity
that does possess a Kac-Moody symmetry and it has been shown that if this
was the case then this symmetry would have to contain a rank eleven
algebra which was called $E_{11}$ \cite{We01}.
It was also argued that $E_{11}$ is  a symmetry of IIA \cite{We01} and IIB
\cite{SchnWe01} supergravity.
Substantial fragments \cite{We01} of  $E_{11}$ were shown to be
symmetries of these  supergravity theories and other evidence has been
given \cite{GaWe02,EnHoTaWe03,EnHoWe03}.
Using similar arguments, it has been proposed that the effective action
of the closed bosonic string generalises in twenty-six dimensions and has
an
associated Kac-Moody algebra of rank twenty-seven called $K_{27}$
\cite{We01} and gravity in $d$ space-time dimensions has an  associated
Kac-Moody algebra of rank  $d$ \cite{LaWe01} respectively 

The Kac-Moody algebras which arose in this
way were of a  special type.
Given a finite-dimensional semi-simple Lie algebra $\lag$ of rank $r$, it
is
well-known how to construct its (non-twisted)
affine and over-extended extensions which
are denoted
$\lag^+$ and $\lag^{++}$
\cite{GoOl85} and are of rank $r+1$ and $r+2$ respectively. The
very-extended
algebra \cite{GaOlWe02} based on $\lag$ is denoted by $\lag^{+++}$
and it has rank
$r+3$. Its Dynkin diagram is found from the Dynkin diagram of the
over-extended algebra $\lag^{++}$ by adding one further node
which is attached to
the  over-extended node by a single line.
The Kac-Moody algebras mentioned just above in
the context of eleven-dimensional supergravity, the effective action of
the closed bosonic string generalised to $d$ space-time dimensions and
gravity in $d$ dimensions are $E_8^{+++}$,  $D_{d-2}^{+++}$, and
$A_{d-3}^{+++}$  respectively.

It is natural to suppose that some similar results exist for any
very-extended Kac-Moody algebra $\lag^{+++}$. For each such algebra
$\lag^{+++}$, one can identify an associated  theory,  denoted in this
paper by ${\cal O}_\lag$,
 consisting of gravity, possible dilatons and forms with
precise couplings  and it has been found by examining Weyl transformations
\cite{EnHoTaWe03} and brane configurations \cite{EnHoWe03}
that these theories show
evidence of very-extended Kac-Moody structures.

The general approach in \cite{We01}  was 
taken up in reference \cite{DaHeNi02} which  considered
eleven-dimensional supergravity as a non-linear realisation of  the
$E_{10}$  subalgebra of
$E_{11}$ in the small tension limit which played a crucial role in  
the work of
references \cite{cosmbill}. The relationship between del
Pezzo surfaces and Borcherds algebras was explored in
\cite{He-LaJuPa02}.

Unlike
finite-dimensional semi-simple Lie algebras
 and affine Lie algebras very little is known about more general
Kac-Moody algebras. In particular, such an elementary property as
all the root
multiplicities  are unknown for
any Kac-Moody algebra that is not of the first two types. It might be
hoped that the there exists a larger  sub-class of Kac-Moody algebras
which are more amenable to study and that this class may contain the
algebras relevant to supergravity and string theories,  namely the
very-extended algebras.

In this paper, we calculate the representation content of very-extended
$\lag^{+++}$ in terms of certain preferred $A$ subalgebras at low levels.
This allows us to predict the field content at low levels of the
corresponding non-linear realisation of $\lag^{+++}$ and we find that this
is in precise agreement with those of the anticipated theory, namely
${\cal O}_\lag$, if the levels are considered to a well-defined
cut-off given by the affine root. The appendices contain details on
some of the next levels which tentatively correspond to new fields.

\end{section}
%
%

\begin{section}{Very-extended Kac-Moody algebras, decompositions and
    the field content}
\label{dectechnique}

\begin{subsection}{Kac-Moody algebras and the process of very-extending}

We follow the conventions of Kac' book \cite{Ka90}. In particular, we
define a Kac-Moody algebra $\lag$ of rank $n$
via its symmetrisable Cartan matrix
$A=(A_{ij})_{i,j=1}^n$
which is related to the simple roots $\alpha_i$ by
\be
A_{ij}=2\frac{(\alpha_i|\alpha_j)}{(\alpha_i|\alpha_i)}.
\ee
The Cartan subalgebra $\csa$ is
generated by the elements
$h_i$ (simple co-roots)
which satisfy
\be
(h_i|h_j)=2\frac{A_{ij}}{(\alpha_i|\alpha_j)}.
\ee
(We will be dealing with non-degenerate $A$ only and thus the
$h_i$ span the Cartan subalgebra $\csa$.)
The Kac-Moody algebra $\lag$
is then generated
as a complex Lie algebra by the elements
$h_i, e_i, f_i$ subject to the relations
\be
   \left[e_i,f_j\right]&=&\delta_{ij}h_i,\\
   \left[h_i,e_j\right]&=&A_{ij}e_j,\\
   \left[h_i,f_j\right]&=&-A_{ij}f_j,\\
   \left[h_i,h_j\right]&=&0,\\
   (\mbox{ad}\,e_i)^{1-A_{ij}}e_j&=&0,\\
   (\mbox{ad}\,f_i)^{1-A_{ij}}f_j&=&0.
\ee
for all $i,j=1,...,n$. The last two relations are called the Serre
relations.
It is a basic fact that the
form $(\,|\,)$ introduced above is symmetric and non-degenerate on
$\csa$. Moreover, it
extends to the invariant symmetric form on the whole Kac-Moody
algebra.

The characteristics of $\lag$ are very different depending on the
properties of the Cartan matrix $A$.
If $A$ is positive definite then the resulting Lie
algebra is finite-dimensional and one from the famous list of
Cartan's classification of
semi-simple Lie algebras (assuming $A$ indecomposable,
i.e. the Dynkin diagram is connected). If $A$ is positive
semi-definite and has precisely one zero eigenvalue the resulting
algebra is an affine algebra. As explained in the introduction our
focus are the so-called very-extended algebras \cite{GaOlWe02} and
their Cartan matrix has signature $(-++\ldots)$.
We will consider the different algebras in the following sections
case by case and list the
relevant Dynkin diagrams there. We will always draw the
very-extended node to the far left and on its right the over-extended
and affine nodes.

We now give some details about the root systems of Kac-Moody algebras
which will be relevant to our analysis. The simple roots $\alpha_i$
are elements
of the dual space of the Cartan subalgebra $\csa$ and so they
naturally act on elements of $\csa$. We denote this action by
$\alpha(h)$ for $\alpha\in\csa^*$ and $h\in\csa$. For example,
we have $\alpha_i(h_j)=A_{ji}$ (which can also be used as a defining
relation). An element
$\alpha=\sum m_i \alpha_i$ of the root lattice $Q$ is called a
root if the corresponding root space
\be
   \lag_\alpha=\left\{x\in\lag\,|\,[h,x]=\alpha(h)x\,\,\mbox{for all
   }h\in\csa\right\}
\ee
is non-trivial and $\alpha\ne 0$. The dimension of the root space is
called the
multiplicity $\mbox{mult}(\alpha)$ of the root.
We call $ht(\alpha)=\sum_{i=1}^n
m_i$ the height of the root and $\alpha^2=(\alpha|\alpha)$ the norm
squared of the root. Kac-Moody algebras possess a triangular decomposition
$\lag=\lap_-\oplus\csa\oplus\lap_+$ and every root is either a sum of
simple roots, and then called a positive root, or the negative thereof.
The positive and negative parts of $\lag$ are exchanged by the
Chevalley involution $\omega$ which acts by
\ben
\label{chevinv}
\omega(h)=-h,\,\,\omega(e_i)=-f_i,\,\,\omega(f_i)=-e_i.
\een

Besides the root lattice $Q$ there is also the weight lattice $P=Q^*$,
which is spanned by the
fundamental weights $\lambda_i$ defined by
$\lambda_i(h_j)=-\delta_{ij}$.\footnote{We employ a convention for the
  definition of the fundamental weights which is suitable for
  hyperbolic and very-extended Kac-Moody algebras. As the inverse
  Cartan matrix of a hyperbolic algebra has only negative entries the
  fundamental weights will lie in the negative half of the root
  lattice with the standard convention $\lambda_i(h_j)=\delta_{ij}$.
  The virtue of the extra sign is that the fundamental weights will
  point in the positive direction. This also has the technical effect of
  interchanging highest weight modules with lowest weight modules but
  we will be cavalier and still use the terminology suitable for
  highest weight modules for clarity and keep this detail in mind.}
We can express any element $\alpha\in Q$ in either basis as
\be
   \alpha=\sum_{i=1}^n m_i \alpha_i
\ee
or as
\be
\alpha=\sum_{i=1}^n p_i \lambda_i.
\ee
We will often write $\alpha$ in terms of its components with respect
to the two bases as
\ben
\label{rootlabels}
   \alpha=(m_1,m_2,\dots,m_n)=[p_1,p_2,\dots,p_n],
\een
using different brackets to indicate the different bases.
The $p_i$ are referred to as Dynkin labels and we can convert from one
set of labels to the other one by using the Cartan matrix. Explicitly,
\ben
\label{basischange}
p_i=-A_{ij}m_j.
\een
As $A$ has integer entries and also $m_i\in\ints$ it is obvious that
also $p_i\in\ints$ for all $i=1,\ldots,n$.

One of the first questions to be asked about a Kac-Moody algebra is
which elements $\alpha$ of the root lattice $Q$ are actually roots of
the algebra. The answer to this question is most easily phrased using
the Weyl group $\cal{W}$
which is generated by the reflections in the simple roots.
$\cal W$ leaves the inner product and the multiplicity invariant.
If the metric on $\csa$ is not definite, like in our case where we
have a Lorentzian space, a potential
root $\alpha$ can have either $\alpha^2>0$
or $\alpha^2\le 0$. In the first case the root is called real and in
the latter imaginary and the set of roots $\Delta$ splits as
$\Delta=\Delta^{re}\uplus\Delta^{im}$.
As the Weyl group leaves the inner product
invariant the sets of real and imaginary roots are closed under the
action of the Weyl group. In fact, all real roots are images of the
simple roots $\alpha_i$ under the Weyl group and all imaginary roots
can be obtained by acting with a Weyl transformation on an element in
the fundamental (Weyl) chamber $\cal{C}$ or in $-\cal{C}$ if that
element has connected support on the Dynkin diagram \cite{Ka90},
i.e. the nodes $i$ for which $m_i\ne 0$ form a connected subgraph of
the Dynkin diagram.
The fundamental
chamber $\cal{C}$ is defined at the subset of $\csa^*$ with only
non-negative Dynkin labels, {\sl i.e.} ${\cal C}=\left\{ \sum
p_i\lambda_i: p_i\ge 0 \,\mbox{for all $i$}\right\}$.

Having established that a certain element $\alpha$ is a root, its
multiplicity can be deduced from the Weyl-Kac
character formula for the trivial representation \cite{Ka90}. Elements
$\alpha\in Q$ which are not roots are assigned vanishing multiplicity
by this formula.
A more useful formulation of the denominator identity is
the Peterson formula \cite{Ka90} and this is the one used here for
performing the calculations in the remainder of the paper. Similarly,
if one is interested in weight multiplicities in (integrable) highest
weight representations, one can use the Freudenthal
formula \cite{Ka90}.

The explicit structure of a Kac-Moody
algebra involving all root multiplicities and structure constants is not
known except for finite-dimensional and affine cases. As we are
interested in the indefinite case we have to rely on recursive methods
based on the Peterson and Freudenthal formulae to
uncover parts of the infinite-dimensional algebra, usually
up to a certain height. We note that there is a generalisation of
Kac-Moody algebras introduced by Borcherds \cite{Bo88} and in some
cases the structure of these generalised Kac-Moody algebras is better
understood in the indefinite cases.

\end{subsection}

\begin{subsection}{Decompositions with respect to a regular subalgebra}
\label{regdec}

We will be interested in decomposing
a given (very-extended) algebra with respect to one of
its regular, finite-dimensional subalgebras. A regular
subalgebra $\las$ here
is given by a Kac-Moody algebra whose Cartan matrix is a
principal proper submatrix of the original Cartan matrix $A$. The Cartan
matrix of the subalgebra will be called $A_{sub}$ in order to distinguish
it from the original one. Let us assume that $A$ has dimension $n$ and
we delete the rows and columns given by $i_1,\ldots,i_g$, then the
subalgebra will have rank $n-g$. We denote the set of indices
belonging to the subalgebra by
$J=\{1,\ldots,n\}\backslash\{i_1,\ldots,i_g\}$, and by $I=\{1,\ldots,n\}$
we denote the set of indices belonging to the full algebra.

As $\las$ acts
upon $\lag$ via the adjoint action, and due to the regularity of the
subalgebra and the triangular decomposition of $\lag$ we can decompose
$\lag$ into a sum of highest weight modules of
$\las$.\footnote{If $\las$ is not
finite-dimensional then there will be one piece which is the adjoint
representation of $\las$ but not of highest weight type, see
\cite{Kl03} for an example.}

We now derive the
conditions on a root $\alpha$ to be a highest weight under the action
of the subalgebra.
This will be a
straight-forward generalisation
of the analysis of \cite{DaHeNi02,We03a,NiFi03}.
We focus on the decomposition of $\lap_+$ since the other half can be
obtained by using the Chevalley involution. According to
(\ref{basischange}), an element $\alpha\in Q$ has Dynkin labels
$[p_1,\ldots,p_n]$ under the action of $\lag$. As the fundamental
weights corresponding to nodes which do not form the Dynkin diagram of
the subalgebra $\las$ are orthogonal to the simple roots of $\las$,
the Dynkin labels on these nodes are unchanged under the action of
$\las$. The condition for such an element to be a 
highest weight element under the action of $\las$ is thus simply
that $p_j\ge 0$ for $j\in J$.

We call the components of an element $\alpha$ on the deleted nodes
the {\sl level}
${\mathbf l}=(l_1,\ldots,l_g)$, so that in components $l_j=m_{i_j}$.
The importance of the level $\mathbf l$ is that it provides a
grading of the algebra $\lag=\oplus_{\mathbf l}\lag_{\mathbf l}$ and
is left invariant under the action of the regular subalgebra.
If we fix the level of an element $\alpha$ to be given by $\mathbf l$
then the values
$p_s$ on the undeleted nodes ($s\in J$) are obtained
from the coefficients of the element $\alpha$ as in
(\ref{basischange}) by
\be
   p_s=-\sum_{i\in I} A_{si}m_i=-\sum_{t\in J}(A_{sub})_{st} m_t -
   \sum_{j=1}^g A_{si_j}l_j.
\ee
Inverting this relation gives
\ben
\label{invertedlabel}
   m_s=-\sum_{t\in J}(A^{-1}_{sub})_{st} p_t -\sum_{t\in
     J}(A^{-1}_{sub})_{st} \sum_{j=1}^g A_{ti_j}l_j
\een
and so we can express a root $\alpha$ at a given level $\mathbf l$ either
by
supplying the Dynkin labels $p_s$ on the remaining nodes belonging to
the subalgebra or the simple
root basis components $m_s$ and the change is determined by
(\ref{invertedlabel}).

The crucial point is that for a finite-dimensional subalgebra the
matrix $A^{-1}_{sub}$ has only positive entries and as $l_j\ge 0$ for
roots and
$A_{ti_j}\le 0$ by the properties of Cartan matrices, it can be seen that
the second term in (\ref{invertedlabel}) is a non-negative number (for
each
$s$). If $\alpha$ is to be a positive root and a highest weight then
both $p_s$ and $m_s$ have to be non-negative integers. Thus there is
only a finite number of allowed solutions to this diophantine problem
at each (fixed) level. Additional constraints derive from the fact
that $\alpha$ has to be a root of $\lag$, the simplest one being that
its norm squared should not exceed the one of the longest root in $\lag$.
For the hyperbolic case this condition is sufficient, but for the
very-extended case it is not. We will refer to solutions of
(\ref{invertedlabel}) with non-negative $m_s$ and $p_s$ that belong to
allowed roots as allowed highest weights.
In the following sections we will carry out this analysis for all
very-extended algebras and for low levels.
Our tables list all the allowed weights, and in order to determine them
one does not need more information about $\lag$ than which roots
appear but not their multiplicities.

To determine the actual outer multiplicity $\mu$ (the number of
copies) with which a given
representation generated by a certain root appears, we use the
character formula. We will see that certain representations that are
possible are actually absent.

There is a nice construction due to Feingold and
Frenkel \cite{FeFr83,FeFrRi93} (see also \cite{KaMoWa88,Kl03})
which allows one to abstractly characterise the decomposition
of $\lag$ into representations of a regular co-rank $1$ subalgebra
at all levels. The coefficient of the node which has been removed will
be referred to as $l$ (without index since $j$ can only take one value
in this case). The representation $U$ that can occur at
first level is
generated by the simple root corresponding to the node which was
deleted. In particular, the corresponding highest weight will be just
the weighted sum of all the fundamental weights of the nodes to which
the deleted node was connected. This is obvious from
(\ref{basischange}) because the simple root of $\lag$ corresponding to
the deleted node has just Dynkin labels equal to minus its column of
the Cartan matrix. Its weight under the action of $\las$ is thus
simply that column with the entry stemming from the diagonal removed.
Most elements obtained by (free)
commutation of the elements at $l=1$ will be present for $l\ge 2$ but not
all of them are elements of $\lag$ since the Serre relations might
remove some of them. The corresponding relations between the node
which was deleted and its neighbours will generate a (not necessarily
irreducible) representation $V$ at $l\ge 2$. The Kac-Moody algebra
$\lag$ is just the quotient of the free algebra
generated by $U$ and the ideal generated by $V$ in the free algebra.
This technique will be used to analyse some of the very-extended
algebras. In particular, in appendix \ref{apppl2} we will present a
general result for the $A$ series at $l=2$ with respect to its maximal
$A$ subalgebra which to the best of our knowledge provides new information
about the structure of an infinite family of indefinite Kac-Moody
algebras.

We will also face the problem of decomposing very-extended algebras
with respect to regular subalgebras of co-rank greater than one. This
can be thought of as successively decomposing with respect to co-rank
one subalgebras, where the order in which these multiple expansions are
done is not important.

In principle, the above considerations completely
determine the decomposition for any specific case. However, the general
answer is still beyond reach as $\lag^{+++}$
is indefinite.

\end{subsection}

\begin{subsection}{Bosonic field content of non-linear realisations of
    very-extended
Kac-Moody algebras}
\label{fieldcont}

A non-linear realisation of a group $G$ with respect to a subgroup $H$
considers an arbitrary element $g$ of the group $G$ which depends on
space-time and constructs
equations of motion, or an action, which are invariant under
 the transformation
$g\to g_0gh$ where
$g_0$ is an arbitrary space-time independent element of $G$ and $h$ is an
element of
$H$ which also depends on space-time. One of the most frequently used
methods to find the invariant theory is to use the Cartan forms
$g^{-1}d g$ which are Lie algebra valued.
These are invariant under the $g_0$ transformations and so
one only has to build quantities out of the Cartan forms which are
invariant under the local $h$ transformations.

For the very-extended algebras we wish to
consider as the basis for our non-linear realisations we take the
subalgebra to be the one which is invariant under the Chevalley involution
of equation (\ref{chevinv}). So we are constructing the non-linear
realisation of $G^{+++}/K(G^{+++})$ where $G^{+++}$ denotes the formal
exponentiation of $\lag^{+++}$ and $K(G^{+++})$ the exponential of the
(maximally compact) subalgebra invariant under the Chevalley
involution.
In this case we may use this local subgroup to  choose
our group elements to be of the form
\ben
\label{iwasawa}
g=\exp(q^a H_a )\exp (\sum_{\alpha\in\Delta_+}\phi_\alpha E_\alpha).
\een
In this equation, $H_a$ are the Cartan subalgebra generators,
$E_\alpha$ are the generators corresponding to a positive root
$\alpha$ and the sum is over all positive roots of the very-extended
algebra. The coefficients $q^a$ and 
$\phi^\alpha$ depend on space-time and are the
fields that appear in the invariant theory that is the non-linear
realisation. Hence, for every positive root of the very-extended algebra
we find a field in the non-linear realisation. The field
associated with a given positive root will have multiple components
if the multiplicity of the root is greater than $1$.

If we are dealing with an internal symmetry then $G$ does not contain
any space-time associated generators and we just let the elements of
$G$ depend on our chosen space-time, as we have done in equation
(\ref{iwasawa}). However, if $G$ does contain space-time generators
these are used to introduce space-time into the group element in a
natural way. In particular, if the algebra contains the translations
$P_a$ then the group element will contain the factor
$\exp(x^aP_a)$. Such a factor was introduced in reference \cite{We00}
which formulated eleven-dimensional supergravity as a non-linear
realisation and was implicit in later discussion involving $E_{11}$.
However, only recently was the $E_{11}$ extension of the space-time
translations discussed and found to include the central charges of the
eleven-dimensional supersymmetry algebra as well as an infinite number of
other objects \cite{We03}. In this paper we will not consider the r\^ole
of
space-time associated generators, although it is expected that they will
play an important r\^ole in a full treatment.

The very-extended algebras contain a preferred regular
$\mathfrak{gl}(d)$, or $A_{d-1}\otimes
\reals$, subalgebra whose generators are denoted by $ K^a{}_b$, the
$\reals$ factor being $\sum_a K^a{}_a$. The space-time generators
$P_a$ are chosen to belong to the vector representation of the
preferred $A_{d-1}$. The group element then contains the  corresponding
factor
$exp\,(x^a P_a)exp\,(h_a{}^b K^a{}_b)$ which in the non-linear
realisation leads to the gravity sector of the theory. Indeed,  one
finds from the non-linear realisation that the
vielbein is given by $e_\mu{}^b=(exp (h))_\mu{}^b$. Hence, the preferred
$A_{d-1}$ controls the space-time sector of the theory.
The $A_{d-1}$ subalgebra can be identified by deleting nodes from
the Dynkin diagram of the very-extended algebra and it always contains the
very-extended,  over-extended and affine  nodes. The $A_{d-1}$
subalgebra is constructed by starting at the very-extended node and
then following the line of long roots so that the rank $d-1$ is maximal,
but
one can also consider submaximal choices.
As we shall see, for a given Dynkin diagram there can sometimes
be more than one choice corresponding to the occurrence of bifurcations in
the diagram. We refer to the part of the Dynkin diagram whose dots
correspond to the preferred $A_{d-1}$ subalgebra  as the gravity line.
The additional factor $\reals$ is part of the Cartan subalgebra
of the very-extended algebra in a way which must also be specified. Having
identified the preferred
$A_{D-1}$ subalgebra we can decompose the field content into
representations of this subalgebra. The advantage of this
decomposition is that the resulting fields can be recognised in terms
that one  is familiar with.

We illustrate the above discussion for the case of
very-extended $E_8$ which we denote as $E_8^{+++}$; this  algebra has
also  been called $E_{11}$ in the literature. It has been conjectured
\cite{We01} that
$E_8^{+++}$ is a symmetry of an extended version of the bosonic
sector of eleven-dimensional supergravity.
The Dynkin diagram of $E_8^{+++}$ is given
in figure \ref{e8pppdynk} where we indicate the regular, largest
possible $A_{10}$ subalgebra which is the preferred subalgebra
relevant to this case. We note that the Cartan subalgebra of
$E_8^{+++}$  is contained entirely in $\mathfrak{gl}(11)$ and indeed can
be
constructed from linear combinations of the generators $K^a{}_a$.

\begin{figure}
\begin{center}
\includegraphics{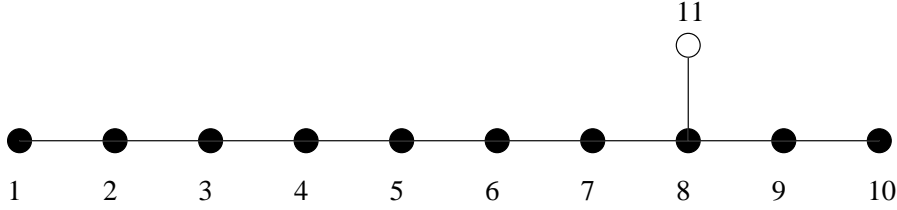}
\caption{\label{e8pppdynk}The Dynkin diagram of $E_8^{+++}$. The
canonical gravity line corresponds to the solid nodes. The Dynkin
diagram of $E_8$ is given by the nodes labelled $4,\ldots,11$, the
affine node is $3$, the over-extended one is $2$ and the very-extended
node is labelled by $1$.}
\end{center}
\end{figure}

Apart from the $K^a{}_b$ generators of $GL(11)$,   the $E_8^{+++}$
algebra contains the positive root generators \cite{We01}
\ben
\label{e8gens}
R^{a_1a_2a_3},\ R^{a_1\dots a_6},\  R^{a_1\dots a_8,b},\dots
\een
which transform as their indices suggest under $\mathfrak{sl}(11)$.
The level of a given root of $E_8^{+++}$
 is just the
number of  times the simple root $\alpha_{11}$ occurs in its
decomposition
into simple roots, as explained in section \ref{regdec}. 
The algebra at
level zero is just
the algebra found by deleting node $11$ from the Dynkin diagram and it is
just the preferred  $A_{10}\times\reals$
subalgebra. The generators in equation
(\ref{e8gens}) are of level $1,2,3$ respectively.

When  constructing the non-linear realisation of
$E_8^{+++}$ we find space-time fields for every positive root of
$E_8^{+++}$. Hence, in addition to the gravity fields $h_a{}^b$ at level
zero we find the fields \cite{We01}
\be
A_{a_1a_2a_3},\ A_{a_1\dots a_6},\  h_{a_1\dots a_8,b} ,\dots
\ee
at levels $1,2,3$ respectively. As already mentioned, the field $h_a{}^b$
 leads to gravity whose vielbein is given by
 $e_\mu{}^a=(exp (h))_\mu{}^a$. The fields at levels one and two
are the third rank gauge field and its dual gauge field respectively in
eleven-dimensional supergravity. Thus in the non-linear realisation
the bosonic non-gravitational fields of eleven-dimensional
supergravity are described by a duality symmetric set
of first order field equations. The field at level
three in equation (\ref{e8gens})
is required to formulate gravity in an analogous
first order system \cite{We01},
although the precise way this occurs
in the interacting theory is unknown. This required formulation
differs from the
one discussed in a five-dimensiona; context 
in \cite{Cu85,Hu00}. Nonetheless, in all theories we
will treat the dual graviton field as part of the gravity sector.
The fields at higher levels are required
in the complete $E_8^{+++}$ invariant theory,
but their precise r\^ole remains to be clarified.
We conclude that the correct field content of the bosonic sector of $11d$
supergravity is encoded in the Cartan subalgebra and the positive root
part
of $E_8^{+++}$ up to level $3$.

Given any very-extended Kac-Moody algebra $\lag^{+++}$ one can construct
its
non-linear realisation, with respect to  a   subalgebra which we take
to be the Cartan involution invariant subalgebra.  
 This theory,
which we denote  by ${\cal V}_\lag$, contains  an infinite number of
fields.
Given our current knowledge of Kac-Moody algebras, this theory can only be
constructed at low levels. As explained above, corresponding to a each
preferred embedding of GL(D) one finds a different effective theory,
however, these are all related by relabelling  of the generators of 
 $\lag^{+++}$ and so are closely related theories. We should in principle 
put a label on ${\cal V}_\lag$ to say which embedding is being considered
but we will refrain from this in this paper and leave the ambiguity as
being understood when it is discussed. 
For example, for the algebra $E_{8}^{+++}$, as
we
just discussed above,  the resulting non-linear realisation containing all
fields up to the level of the affine root  is just eleven-dimensional
supergravity if we take the obvious GL(11) embedding. Precisely
how the fields corresponding to higher levels of
$E_{8}^{+++}$ modify this theory is at present unknown. 

Clearly, if the theory ${\cal V}_\lag$ is
dimensionally reduced on a torus then the resulting theory will contain
scalars that will belong to a non-linear realisation of a sub-algebra of
the very-extended algebra  which is preserved by the  torus.
In the reduction to three dimensions all
the fields can be dualised to scalars and so one finds a theory consisting
of scalars alone that is a non-linear realisation of some group $G$ with
respect to its maximally compact subgroup. In fact, it is known that one
can obtain all  of the finite-dimensional semi-simple Lie group in this
way
\cite{CrJuLuePo99}
by starting with an appropriate theory. The unreduced theory
with the maximal space-time dimension from which one can start to obtain
the
non-linear realisation of $G$ in three dimensions has become known as the
maximally oxidised theory and we denote it by ${\cal O}_\lag$.
The most well-known
example is eleven-dimensional supergravity which is the maximally
oxidised theory of the three dimensional theory whose scalars belong to
the  non-linear realisation of $E_8$ with respect to $D_8$.
Eleven-dimensional supergravity is thus called
${\cal O}_{E_8}$ in this language.
We note  that in general
the dimensional reduction of a theory that contains gravity, dilatons and
gauge forms does not lead to scalars that form a non-linear realisation
and indeed only occurs   for a restricted set of theories each of
which has a specified   field content and a precise set of couplings
between the fields \cite{LaWe01}.

One might anticipate, in view of the discussion of reduction,
that the non-linear realisation of $\lag^{+++}$,
i.e.
${\cal V}_\lag$, will contain  up to a certain low level the theory
${\cal O}_\lag$.
As ${\cal O}_\lag$ is characterised by $\lag$ it is natural to look
at the fields generated by the $\lag$ subalgebra of $\lag^{+++}$ and
this will provide a cut-off criterion in our decomposition.
In this paper we calculate the generators and corresponding field content
of
$\lag^{+++}$  in terms of  representations of the preferred $A_{d-1}$
subalgebra at low levels. We find that  the corresponding field content
up to (and including) the lowest level at  which an affine root occurs is in
one-to-one 
correspondence with  the bosonic
field content of the theory which is associated with ${\cal O}_\lag$.
This provides a significant check on the conjecture that extensions
of the ${\cal O}_\lag$ theories exist and do possess a
non-linearly realised
$G^{+++}$ symmetry.

\end{subsection}

\end{section}

%
%

\begin{section}{Simply-laced cases}

In this section we begin by analysing the representation content of
the very-extension $\lag^{+++}$ of simply-laced algebras $\lag$ with
respect
to some of their regular $A_{d-1}$ subalgebras. The low-lying
representations correspond to the bosonic field content of various
known theories containing gravity, forms and dilatons.
As mentioned in section \ref{fieldcont},
low-lying here refers to precisely those
representations whose level does not exceed the level at which the
representation corresponding to the first affine root (of the
non-twisted affine algebra obtained in the process of very-extension)
occurs. For the $E$ series this is the same as considering all
representations generated up to the height of the first affine root.
Then the relevant roots of the very-extended algebras all belong to the
affinised finite-dimensional Lie algebra that has been
extended. Actually, with the exception of the representations dual to
scalar fields coming from the Cartan subalgebra, 
they all come from the finite-dimensional version
of the algebra. However,
in the corresponding physical theory Lorentz covariance with
respect to the whole $\mathfrak{sl}(d)$ gravity algebra implies the
occurence of the very-extended version of those algebras. The
finite-dimensional
Lie subalgebras correspond precisely to the known symmetries of
the scalar cosets after compactification to three dimensions. In
this way, affinisation, over- and very-extension can tentatively
be understood as
coset symmetries of the arising scalars in 2, 1, 0 dimensions.
Some representations of the very-extended simply-laced algebras for
levels greater than that of the first affine root are listed in the
appendices.

\begin{subsection}{$E$ series}

\begin{subsubsection}{$E_8^{+++}$, the bosonic part of $11d$
supergravity and type {\rm IIA} and {\rm IIB} theory}
\label{e8ppp}

We summarise the discussion of $E_8^{+++}$ with respect to its
$A_{10}$ subalgebra given in section \ref{fieldcont}
in table \ref{e8pppdec}.

\begin{table}
\begin{tabular}{cccrrrr}
$l$&$A_{10}$ weight&$E_8^{+++}$ element
$\alpha$&$\alpha^2$&$ht(\alpha)$&$\mu$&interpretation\\
\hline
0&[1,0,0,0,0,0,0,0,0,1]&(1,1,1,1,1,1,1,1,1,1,0)&2&10&1&$h_a{}^b$\\
1&[0,0,0,0,0,0,0,1,0,0]&(0,0,0,0,0,0,0,0,0,0,1)&2&1&1&$A_{(3)}$\\
2&[0,0,0,0,1,0,0,0,0,0]&(0,0,0,0,0,1,2,3,2,1,2)&2&11&1&$\tilde{A}_{(6)}$\\
3&[0,0,1,0,0,0,0,0,0,1]&(0,0,0,1,2,3,4,5,3,1,3)&2&22&1&$\tilde{A}_{(8,1)}$\\
3&[0,1,0,0,0,0,0,0,0,0]&(0,0,1,2,3,4,5,6,4,2,3)&0&30&0&$\tilde{A}_{(9)}^\phi$
\end{tabular}
\caption{\label{e8pppdec}The first three levels of $E_8^{+++}$
decomposed with respect to its regular $A_{10}$ subalgebra.}
\end{table}

We have listed all representations that are possible up to height
$30$, however the last column $\mu$ contains the outer multiplicity with
which the given representation actually occurs. If the outer multiplicity
is zero, then the Serre relations forbid the occurrence of this
representation. The table is complete for the first three levels and
all representations which occur at higher levels will be generated
by $E_8^{+++}$ elements of height greater than $30$.

It was suggested in \cite{We01,We03a} that this representation content
can be associated with the bosonic field contents of eleven-dimensional
supergravity in an appropriate formulation. Thereby, the generators of
$GL(11)$ contain the degrees of freedom of the
vielbein $e_\mu{}^a$. The level $l=1$ representation is the antisymmetric
rank three tensor under $A_{10}$ and corresponds to the three-form potential
$A_{(3)}$ in $11d$ supergravity. At level $l=2$ we find a six-form
representation whose field strength is dual to the one of the three-form
potential in $11d$. The field at $l=3$ can be seen as the dual field to
the vielbein, subject to the remarks in section \ref{fieldcont}.
We also see that the nine-form
representation is forbidden by the Serre relation. In eleven-dimensional
supergravity this relates to the fact that there is neither a dilaton
field nor its dual.

The physical interpretation of representations beyond this point has
yet to be given. A list of some of
these representations can be found in
\cite{We03a, NiFi03}.
We conclude that the correct field content of the bosonic sector of
maximal $11d$
supergravity is encoded in the Cartan subalgebra and the positive root
part
of $E_8^{+++}$ up to level $3$
with respect to its $A_{10}$ subalgebra.\\

We now show how to obtain both ten-dimensional maximal supergravities
from $E_8^{+++}$ by decomposing into representations of different
regular $A_9$
subalgebras. As the Dynkin diagram has a bifurcation,
there are actually two different choices for this and the
different $A_9$ subalgebras are depicted in figure
\ref{e8pppabdynk}.\\

\begin{figure}
\begin{center}
\includegraphics{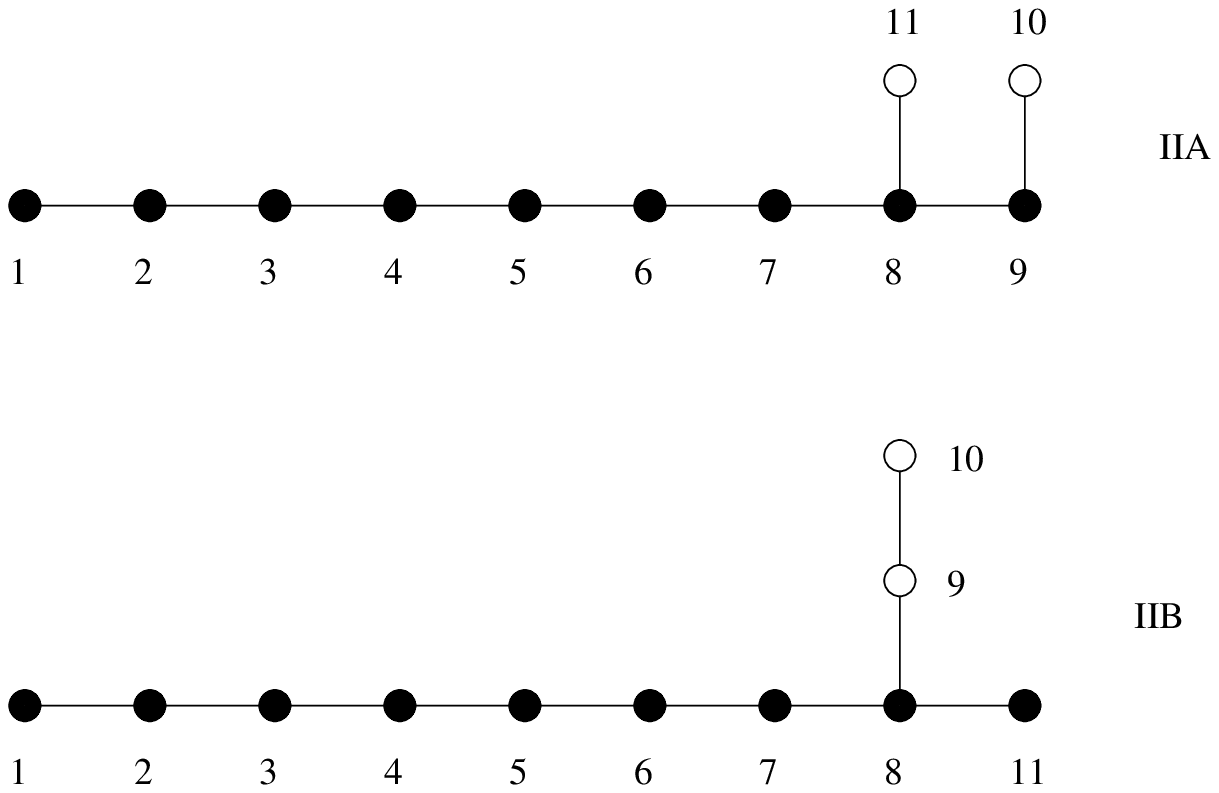}
\caption{\label{e8pppabdynk}The Dynkin diagram of $E_8^{+++}$ drawn
with an $A_9$ subalgebra (solid nodes) to yield a ten-dimensional
theory.}
\end{center}
\end{figure}

{\bf  IIA: }
First, consider the case where the gravity line is chosen as depicted
in the top half of the figure. We will label the decomposition by a pair
of non-negative integers $(l_1,l_2)$ where $l_1$ corresponds to the
entry on the node denoted $11$ in the diagram and $l_2$ to the node
denoted $10$. Doing the decomposition gives the $\mathfrak{sl}(10)$
content
of table \ref{e8pppdeca} on the first few levels.

\begin{table}
\begin{tabular}{cccrrrr}
$(l_1,l_2)$&$A_9$ weight&$E_8^{+++}$ element
$\alpha$&$\alpha^2$&$ht(\alpha)$&$\mu$&Interpretation\\
\hline
(0,0)&[1,0,0,0,0,0,0,0,1]&(1,1,1,1,1,1,1,1,1,0,0)&2&9&1&$h_a{}^b$\\
(0,0)&[0,0,0,0,0,0,0,0,0]&(0,0,0,0,0,0,0,0,0,0,0)&0&0&1&$\phi$\\
(1,0)&[0,0,0,0,0,0,0,1,0]&(0,0,0,0,0,0,0,0,0,0,1)&2&1&1&$B_{(2)}$\\
(0,1)&[0,0,0,0,0,0,0,0,1]&(0,0,0,0,0,0,0,0,0,1,0)&2&1&1&$A_{(1)}$\\
(1,1)&[0,0,0,0,0,0,1,0,0]&(0,0,0,0,0,0,0,1,1,1,1)&2&4&1&$A_{(3)}$\\
(2,1)&[0,0,0,0,1,0,0,0,0]&(0,0,0,0,0,1,2,3,2,1,2)&2&11&1&$\tilde{A}_{(5)}$\\
(2,2)&[0,0,0,1,0,0,0,0,0]&(0,0,0,0,1,2,3,4,3,2,2)&2&17&1&$\tilde{B}_{(6)}$\\
(3,1)&[0,0,1,0,0,0,0,0,0]&(0,0,0,1,2,3,4,5,3,1,3)&2&22&1&$\tilde{A}_{(7)}$\\
(3,2)&[0,0,1,0,0,0,0,0,1]&(0,0,0,1,2,3,4,5,3,2,3)&2&23&1&$\tilde{A}_{(7,1)}$\\
(3,2)&[0,1,0,0,0,0,0,0,0]&(0,0,1,2,3,4,5,6,4,2,3)&0&30&1&$\tilde{A}^\phi_{(8)}$
\end{tabular}
\caption{\label{e8pppdeca}The first few levels of $E_8^{+++}$
decomposed with respect to its regular $A_{9}$ subalgebra obtained by
deleting nodes $11$ and $10$ corresponding to a grading by $l_1$ and
$l_2$ respectively. The representation content is related to type
{\rm IIA} supergravity in the text.}
\end{table}

Looking at the fields,
we find precisely the generators corresponding to the bosonic
part of {\rm IIA} supergravity, namely the NS-NS two-form, the
RR one- and three-forms along with their duals. At level $(0,0)$ the
generators of $\mathfrak{sl}(10)$ and two additional fields
are found. As usual, they
contain the gravitational degrees of freedom as $\mathfrak{gl}(10)$ and
the additional scalar $\phi$ which can be
interpreted as the dilaton field.
Its dual occurs together
with the dual gravity field at level $(3,2)$. The precise relation between
the
field generators in a non-linear realisation of eleven-dimensional
supergravity and ten-dimensional IIA supergravity was already given in
\cite{We01} and can be understood as follows: the positive root component
of the $K^a{}_{11}$ generator that stretches in the eleventh dimension
is interpreted as the Kaluza-Klein vector generator of the dimensionally
reduced theory $K^a{}_{11}\sim R^{10}$. Similarly, the additional
Cartan subalgebra scalar
generator $\phi$
is proportional to the $K^{11}{}_{11}$ part. For the gauge field
generators one finds canonically $R^{ab\,11} \sim R^{ab}$, where the
former is the three-form of eleven-dimensional supergravity, while the
latter is the 2-form generator in ten-dimensional IIA supergravity.\\

{\bf IIB:}
There is another possible regular $A_9$ subalgebra as shown in part
(b) of figure \ref{e8pppabdynk}, and its $\mathfrak{sl}(10)$
representation content is given in table \ref{e8pppdecb} up to the
level of the affine root.

\begin{table}
\begin{tabular}{cccrrrr}
$(l_1,l_2)$&$A_9$ weight&$E_8^{+++}$ element
$\alpha$&$\alpha^2$&$ht(\alpha)$&$\mu$&Interpretation\\
\hline
(0,0)&[1,0,0,0,0,0,0,0,1]&(1,1,1,1,1,1,1,1,0,0,1)&2&9&1&$h_a{}^b$\\
(0,0)&[0,0,0,0,0,0,0,0,0]&(0,0,0,0,0,0,0,0,0,0,0)&0&0&1&$\phi$\\
(1,0)&[0,0,0,0,0,0,0,0,0]&(0,0,0,0,0,0,0,0,0,1,0)&2&1&1&$\chi$\\
(0,1)&[0,0,0,0,0,0,0,1,0]&(0,0,0,0,0,0,0,0,1,0,0)&2&1&1&$B_{(2)}$\\
(1,1)&[0,0,0,0,0,0,0,1,0]&(0,0,0,0,0,0,0,0,1,1,0)&2&2&1&$A_{(2)}$\\
(1,2)&[0,0,0,0,0,1,0,0,0]&(0,0,0,0,0,0,1,2,2,1,1)&2&7&1&$A_{(4)}$\\
(1,3)&[0,0,0,1,0,0,0,0,0]&(0,0,0,0,1,2,3,4,3,2,2)&2&16&1&$\tilde{B}_{(6)}$\\
(2,3)&[0,0,0,1,0,0,0,0,0]&(0,0,0,0,2,2,3,4,3,2,2)&2&17&1&$\tilde{A}_{(6)}$\\
(2,4)&[0,0,1,0,0,0,0,0,1]&(0,0,0,1,2,3,4,5,4,2,2)&2&23&1&$\tilde{A}_{(7,1)}$\\
(1,4)&[0,1,0,0,0,0,0,0,0]&(0,0,1,2,3,4,5,6,4,1,3)&2&29&1&$\tilde{A}^\chi_{(8)}$\\
(2,4)&[0,1,0,0,0,0,0,0,0]&(0,0,1,2,3,4,5,6,4,2,3)&0&30&1&$\tilde{A}^\phi_{(8)}$
\end{tabular}
\caption{\label{e8pppdecb}The first few levels of $E_8^{+++}$
decomposed with respect to its regular $A_{9}$ subalgebra obtained by
deleting nodes $10$ and $9$ corresponding to a grading by $l_1$ and
$l_2$ respectively. The representation content is related to type
{\rm IIB} supergravity in the text.}
\end{table}

The last
column gives the interpretation in terms of fields of IIB supergravity
in ten dimensions. There is again a one-to-one correspondence if we
accept that gauge fields and gravity have to be treated on an equal
footing by introducing duals to either. Indeed, there is a non-linear
realisation of IIB supergravity which employs all those fields
\cite{SchnWe01}. The occurrence of only one four-form potential implies
that its field strength will have to be self-dual.

More details on the higher levels for the {\rm IIA} and {\rm IIB}
decompositions can be found in appendix \ref{e8pppl2}.

\end{subsubsection}

\begin{subsubsection}{$E_7^{+++}$}
\label{e7ppp}

The Dynkin diagram of very-extended $E_7$ is given in figure
\ref{e7pppdynk}. Again, there are two alternative choices for the
gravity subalgebra as the diagram has a bifurcation. As depicted in
figure \ref{e7pppdynk} we can choose either an $A_9$ or an $A_7$
subalgebra corresponding to theories
in $8$ or $10$ dimensions respectively.\\

\begin{figure}
\begin{center}
\includegraphics{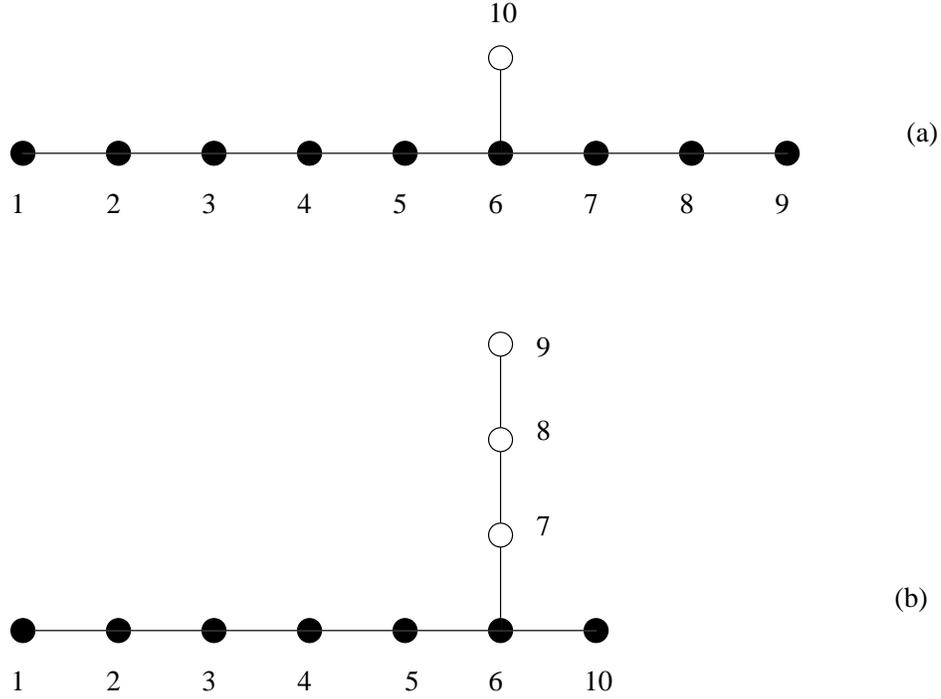}
\caption{\label{e7pppdynk}The Dynkin diagram of $E_7^{+++}$ drawn
with an $A_9$ subalgebra to yield a ten-dimensional
theory (a) or with an $A_7$ subalgebra to obtain an eight-dimensional
theory (b).}
\end{center}
\end{figure}

{\bf (a) Decomposition into representations of $A_9$:}
A similar analysis to above gives the representation content
with respect to $\mathfrak{sl}(10)$ presented in table \ref{e7pppdec10}
where we label the component on the node $10$ by
$l$.

\begin{table}
\begin{tabular}{cccrrrr}
$l$&$A_9$ weight&$E_7^{+++}$ element
$\alpha$&$\alpha^2$&$ht(\alpha)$&$\mu$&interpretation\\
\hline
0&[1,0,0,0,0,0,0,0,1]&(1,1,1,1,1,1,1,1,1,0)&2&9&1&$h_a{}^b$\\
1&[0,0,0,0,0,1,0,0,0]&(0,0,0,0,0,0,0,0,0,1)&2&1&1&$A_{(4)}$\\
2&[0,0,1,0,0,0,0,0,1]&(0,0,0,1,2,3,2,1,0,2)&2&11&1&$A_{(7,1)}$\\
2&[0,1,0,0,0,0,0,0,0]&(0,0,1,2,3,4,3,2,1,2)&0&18&0&$A_{(8)}^\phi$
\end{tabular}
\caption{\label{e7pppdec10}The first three levels of $E_7^{+++}$
decomposed with respect to its regular $A_{9}$ subalgebra.}
\end{table}

All other representations lie above the level where the first
affine root appears. We find a four-form ($l=1$) and the dual
gravity generator ($l=2$) apart from the adjoint gravity generator
at level $l=0$. This theory is a consistent non-supersymmetric
truncation of {\rm IIB}
supergravity as can be easily found out by truncating the algebra called
$g_{IIB}$ in \cite{SchnWe01}. We note that this theory is listed as the
potential oxidation endpoint for the $E_7$ coset in \cite{CrJuLuePo99}.
We also note that a potential dual scalar (dual dilaton) has outer
multiplicity zero ($\mu=0$) and is thus absent (and so is the dilaton
itself). This is analogous to the eleven-dimensional case of $E_8^{+++}$.

The truncation of {\rm IIB} to the content at hand can be understood
purely
algebraically by embedding $E_7^{+++}$ in $E_8^{+++}$. We choose as
simple roots for the $E_7^{+++}$ subalgebra the gravity line and the
element $\alpha_7+2\alpha_8+2\alpha_9+\alpha_{10}+\alpha_{11}$ of
$E_8^{+++}$. From this embedding we see that the
elements of $E_8^{+++}$ which also give rise to $E_7^{+++}$ generators
must obey $2l_1=l_2$ in the type {\rm IIB} table \ref{e8pppdecb}. The
fields satisfying this constraint
are the vielbein $h_a{}^b$, the Cartan subalgebra scalar
$\phi$, a four-form potential $A_{(4)}$, the dual graviton
$K^{a_1\ldots a_7,b}$ and the field $A_{(8)}^\phi$ dual $\phi$. As the
Cartan subalgebra of $E_7^{+++}$ has one dimension less than that of
$E_8^{+++}$ it is clear that $\phi$ is absent in the $E_7^{+++}$
analysis and it is consistent that so is the dual field $A_{(8)}^\phi$.
We also find that a reduction from
{\rm IIA}
or eleven-dimensional supergravity is not immediately possible because
of the different $A_9$ subalgebra. \\

{\bf (b) Decomposition into representations of $A_7$:}

We can also consider embedding an $A_7$ subalgebra as indicated in
part (b) of figure \ref{e7pppdynk} where we expect to find an
eight-dimensional theory. From the embedding $E_7^{+++}\subset
E_8^{+++}$ we expect this to be a truncation of maximal $8d$
supergravity, not necessarily supersymmetric.
We list all representations
occurring up to the level of the affine $E_7^{+}$ root in table
\ref{e7pppdec8}.

\begin{table}
\begin{tabular}{cccrrrr}
$(l_1,l_2,l_3)$&$A_7$ weight&$E_7^{+++}$ element
$\alpha$&$\alpha^2$&$ht(\alpha)$&$\mu$&interpretation\\
\hline
(0,0,0)&[1,0,0,0,0,0,1]&(1,1,1,1,1,1,0,0,0,1)&2&7&1&$h_a{}^b$\\
(0,0,0)&[0,0,0,0,0,0,0]&(0,0,0,0,0,0,0,0,0,0)&0&0&2&$\phi$\\
(1,0,0)&[0,0,0,0,0,0,0]&(0,0,0,0,0,0,0,0,1,0)&2&1&1&$\chi$\\
(0,1,0)&[0,0,0,0,0,0,0]&(0,0,0,0,0,0,0,1,0,0)&2&1&1&$\chi$\\
(0,0,1)&[0,0,0,0,0,1,0]&(0,0,0,0,0,0,1,0,0,0)&2&1&1&$A_{(2)}$\\
(1,1,0)&[0,0,0,0,0,0,0]&(0,0,0,0,0,0,0,1,1,0)&2&2&1&$\chi$\\
(0,1,1)&[0,0,0,0,0,1,0]&(0,0,0,0,0,0,1,1,0,0)&2&2&1&$A_{(2)}$\\
(1,1,1)&[0,0,0,0,0,1,0]&(0,0,0,0,0,0,1,1,1,0)&2&3&1&$A_{(2)}$\\
(0,1,2)&[0,0,0,1,0,0,0]&(0,0,0,0,1,2,2,1,0,1)&2&7&1&$\tilde{A}_{(4)}$\\
(1,1,2)&[0,0,0,1,0,0,0]&(0,0,0,0,1,2,2,1,1,1)&2&8&1&$\tilde{A}_{(4)}$\\
(1,2,2)&[0,0,0,1,0,0,0]&(0,0,0,0,1,2,2,2,1,1)&2&9&1&$\tilde{A}_{(4)}$\\
(1,2,3)&[0,0,1,0,0,0,1]&(0,0,0,1,2,3,3,2,1,1)&2&13&1&$\tilde{A}_{(5,1)}$\\
(0,1,3)&[0,1,0,0,0,0,0]&(0,0,1,2,3,4,3,1,0,2)&2&16&1&$\tilde{A}_{(6)}^\chi$\\
(1,1,3)&[0,1,0,0,0,0,0]&(0,0,1,2,3,4,3,1,1,2)&2&17&1&$\tilde{A}_{(6)}^\chi$\\
(0,2,3)&[0,1,0,0,0,0,0]&(0,0,1,2,3,4,3,2,0,2)&2&17&1&$\tilde{A}_{(6)}^\chi$\\
(1,2,3)&[0,1,0,0,0,0,0]&(0,0,1,2,3,4,3,2,1,2)&0&18&2&$\tilde{A}_{(6)}^\phi$
\end{tabular}
\caption{\label{e7pppdec8}The first three levels of $E_7^{+++}$
decomposed with respect to its regular $A_{7}$ subalgebra of figure
\ref{e7pppdynk}. $l_1, l_2$ and $l_3$ correspond to nodes 9, 8 and 7
respectively.}
\end{table}

We find a
triplet of positive root scalars and two Cartan subalgebra scalars,
the vielbein, and the duals of all
these fields (note outer
multiplicity 2 for the dual dilaton). There are also
three two-forms with their dual four-forms. This is the field content
of a non-supersymmetric truncation of maximal ${\cal N}=2$
eight-dimensional supergravity whose bosonic field content comprises a
vielbein, a triplet of two-forms, a triplet of one-forms, one
three-form potential and an additional (axion) scalar. These fields
arise from reduction of the maximal eleven-dimensional theory and so
transform under a global $SL(3)$.
Furthermore the supergravity theory has
scalars in the coset $SL(3)/SO(3)$.

The r\^ole of the global $SL(3)$ can also be seen already in the
Dynkin diagram of $E_7^{+++}$
where we have an $A_2$ which is not attached to the
gravity line, consisting of the nodes labelled $8$ and $9$ in diagram
\ref{e7ppp}, part (b). In the non-linear realisation, the fields are
controlled by the coset of $E_7^{+++}$ by its Chevalley invariant
subalgebra and the coset contains a part belonging to this $A_2$ which
is just $SL(3)/SO(3)$ after exponentiation of the real algebras
and acts on the scalars arising from this part
of the algebra in the decomposition. The coset is parametrised by five
elements and three are associated with positve roots of $E_7^{+++}$
and two with the Cartan subalgebra as is also apparent in the
table. The positive roots of the
$A_2$ subalgebra also act on the generators forming the two-form
representation on level $(0,0,1)$ in the table and so generate another
two two-forms, all of which transform under the $\mathfrak{sl}(3)$.
These fields
together with the vielbein are retained in the non-supersymmetric
truncation of maximal ${\cal N}=2$ eight-dimensional supergravity. Again
the Kac-Moody algebra also contains duals of all the fields precisely
up to the level of the first affine root.

In appendix \ref{e7pppl2} we list details for the higher levels
of a decomposition of $E_7^{+++}$ with respect to its regular $A_8$
gravity subalgebra, corresponding to a nine-dimensional theory.

\end{subsubsection}

\begin{subsubsection}{$E_6^{+++}$}
\label{e6ppp}

This section deals with very-extended $E_6$ which is
denoted by $E_6^{+++}$. The Dynkin diagram is given in figure
\ref{e6pppdynk}. Here there is only one canonical maximal choice
of a gravity subalgebra $A_7$ because the possibilities are exchanged
by a diagram automorphism. The list of resulting representations
of $\mathfrak{sl}(8)$ up to level $(1,2)$ (first affine root) is provided
in table
\ref{e6pppdec}. The level $(l_1,l_2)$ corresponds to the components of a
positive root on the nodes $9$ and $8$ respectively.

\begin{figure}
\begin{center}
\includegraphics{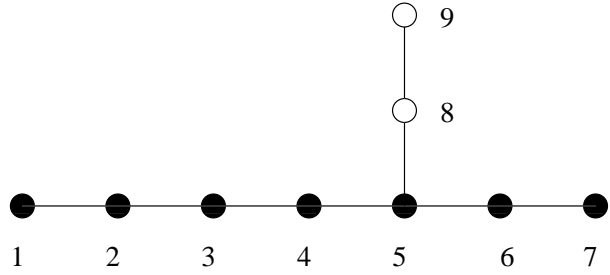}
\caption{\label{e6pppdynk}The Dynkin diagram of $E_6^{+++}$ drawn
with its $A_7$ subalgebra.}
\end{center}
\end{figure}

\begin{table}
\begin{tabular}{cccrrrr}
$(l_1,l_2)$&$A_7$ weight&$E_6^{+++}$ element
$\alpha$&$\alpha^2$&$ht(\alpha)$&$\mu$&Interpretation\\
\hline
(0,0)&[1,0,0,0,0,0,1]&(1,1,1,1,1,1,1,0,0)&2&7&1&$h_a{}^b$\\
(0,0)&[0,0,0,0,0,0,0]&(0,0,0,0,0,0,0,0,0)&0&0&1&$\phi$\\
(1,0)&[0,0,0,0,0,0,0]&(0,0,0,0,0,0,0,0,1)&2&1&1&$\chi$\\
(0,1)&[0,0,0,0,1,0,0]&(0,0,0,0,0,0,0,1,0)&2&1&1&$A_{(3)}$\\
(1,1)&[0,0,0,0,1,0,0]&(0,0,0,0,0,0,0,1,1)&2&2&1&$\tilde{A}_{(3)}$\\
(1,2)&[0,0,1,0,0,0,1]&(0,0,0,1,2,1,0,2,1)&2&7&1&$\tilde{A}_{(5,1)}$\\
(0,2)&[0,1,0,0,0,0,0]&(0,0,1,2,3,2,1,2,0)&2&11&1&$\tilde{A}^\chi_{(6)}$\\
(1,2)&[0,1,0,0,0,0,0]&(0,0,1,2,3,2,1,2,1)&0&12&1&$\tilde{A}^\phi_{(6)}$
\end{tabular}
\caption{\label{e6pppdec}The first few levels of $E_6^{+++}$
decomposed with respect to its regular $A_{7}$ subalgebra obtained by
deleting nodes $9$ and $8$ corresponding to a grading by $l_1$ and
$l_2$ respectively.}
\end{table}

We find a positive root
scalar (axion) and its dual, but also the vielbein and a Cartan
subalgebra scalar (dilaton)
on level $(0,0)$ and their duals. There also are
two three-forms generated by positive roots. The dual
graviton and dual dilaton occur on the same level. This is precisely
the field content of the oxidation endpoint of the supersymmetric
$E_6$ coset theory
described in \cite{CrJuLuePo99} where the four-form field strength and
its dual form a doublet under a global $SL(2,\reals)$.
Details on the higher levels can be found in appendix \ref{e6pppl2}.

\end{subsubsection}

\end{subsection}

\begin{subsection}{$D$ series}
\label{dppp}

The Dynkin diagram of the very-extended $D$ algebras is given in figure
\ref{dpppdynk} and the gravity line is indicated as usual by the solid
nodes. Different from the cases considered so far, there are actually
two nodes in the gravity subalgebra where other subalgebras couple to it.
The gravity line is an $A_{n-2}$ algebra, and we expand around the
nodes marked as $n$ and $n-1$. Expanding the node $n$ leaves a $D_{n-1}$
subalgebra, further expansion around $n-1$ yields $A_{n-2}$. If the
expansion nodes are exchanged then we get a $E_7^{+(n-8)}$
by which we mean an $E_7$ algebra extended $n-8$ times.
This is an indefinite Kac-Moody algebra, and belongs to
the class of algebras considered in \cite{GaOlWe02} but is in general
not very-extended.

\begin{figure}
\begin{center}
\includegraphics{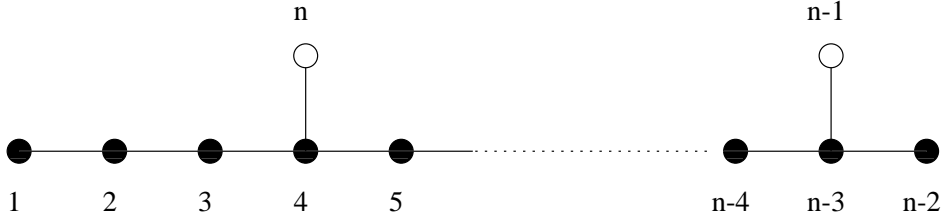}
\caption{\label{dpppdynk}The Dynkin diagram of $D_{n-3}^{+++}$ drawn
with its $A_{n-2}$ subalgebra as gravity given by the solid nodes.}
\end{center}
\end{figure}

According to  the level criterion for the physical spectrum of ${\cal
  O}_\lag$, we are
interested in finding all representations on levels less than $(1,1)$
which is the level of the affine root coming from the affine $D_{n-3}$
subalgebra. It is possible to apply a mild generalisation
of the Feingold-Frenkel method to find the decomposition up to this level
and the result is presented in table
\ref{dpppdec}. This list
is complete since the tensor product of the representations at levels
$(0,1)$ and $(1,0)$ has three summands, one of which lies in the
ideal described in section \ref{regdec}
and thus has to be removed. The other
two are the listed ones. The outer multiplicity of the first one is fixed
by it being generated by the lowest element on that level and being a
real root. The
outer multiplicity of the affine $D$ root can be computed by noting
that its multiplicity in the algebra is $n-3$ and its multiplicity as
a weight in the other representation on that level is $n-4$.

\begin{table}
\begin{tabular}{cccrrrr}
$(l_1,l_2)$&$A_{n-2}$ weight&$D_{n-3}^{+++}$ element
$\alpha$&$\alpha^2$&$ht(\alpha)$&$\mu$&\\
\hline
(0,0)&[1,0,0,0,0,\dots,0,0,1]&(1,1,1,1,\dots,1,1,0,0)&2&$n-2$&1&$h_a{}^b$\\
(0,0)&[0,0,0,0,0,\dots,0,0,0]&(0,0,0,0,\dots,0,0,0,0)&0&0&1&$\phi$\\
(0,1)&[0,0,0,0,0,\dots,0,1,0]&(0,0,0,0,\dots,0,0,1,0)&2&1&1&$A_{(2)}$\\
(1,0)&[0,0,0,1,0,\dots,0,0,0]&(0,0,0,0,\dots,0,0,0,1)&2&1&1&$\tilde{A}_{(n-5)}$\\
(1,1)&[0,0,1,0,0,\dots,0,0,1]&(0,0,0,1,\dots,1,0,1,1)&2&$n-4$&1&$\tilde{A}_{(n-4,1)}$\\
(1,1)&[0,1,0,0,0,\dots,0,0,0]&(0,0,1,2,\dots,2,1,1,1)&0&$2n-8$&1&$\tilde{A}_{(n-3)}^\phi$
\end{tabular}
\caption{\label{dpppdec}The first levels of $D_{n-3}^{+++}$
decomposed with respect to its regular $A_{n-2}$ subalgebra obtained by
deleting nodes $n$ and $n-1$ giving a grading by $l_1$ and $l_2$
respectively.}
\end{table}

The
field content on these levels is thus a vielbein in $n-1$ dimensions,
a scalar coming from the Cartan subalgebra which we interpret as
dilaton and an antisymmetric two-form.
Furthermore we find the duals to all these
fields.
This is just the set of massless states of
closed bosonic string theory in $n-1$ dimensions, in agreement with
the conjecture in
\cite{We01} and also in agreement with the content of the oxidised
theory \cite{CrJuLuePo99}.
We also note that the Dynkin diagram has a bifurcation at
node $4$ and the other choice of gravity $A_5$ subalgebra would lead
to  a theory in six dimensions. The field
content up to the level of the affine root of that theory exists of
$(n-6)(n-7)$ positive root scalars and their duals, $2(n-6)$
2-forms, $n-6$ Cartan subalgebra scalars
and their duals, along with a dual graviton.

In appendix \ref{d8pppl2} we list the higher levels for the particular
case of $D_8^{+++}$.

\end{subsection}

\begin{subsection}{$A$ series}
\label{appp}

In this section we analyse the very-extended algebras of the $A$
series on level one which is the relevant bound for the spectrum as
the first affine root occurs at this level. The Dynkin diagram of
$A_{n-3}^{+++}$ is given in figure \ref{apppdynk}.

\begin{figure}
\begin{center}
\includegraphics{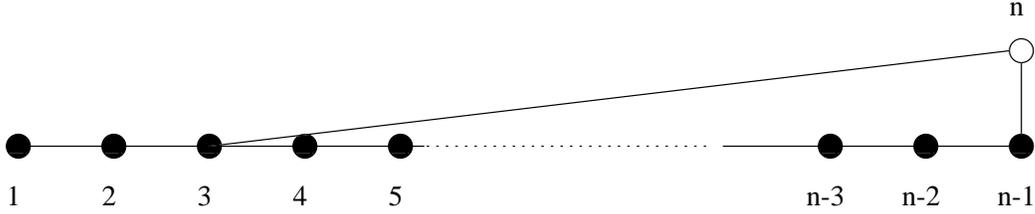}
\caption{\label{apppdynk}The Dynkin diagram of $A_{n-3}^{+++}$ drawn
with its $A_{n-1}$ subalgebra as gravity.}
\end{center}
\end{figure}

The representation content with respect to the indicated $A_{n-1}$
subalgebra was presented in \cite{We03a} based on an analysis of the
possible weights and the Serre relations. Here we deduce the same
results using the Feingold-Frenkel mechanism (see section
\ref{dectechnique}). Table \ref{apppdec} contains the results and all
other representations occur at higher level.

\begin{table}
\begin{tabular}{cccrrrr}
$l$&$A_{n-1}$ weight&$A_{n-3}^{+++}$ element
$\alpha$&$\alpha^2$&$ht(\alpha)$&$\mu$&Interpretation\\
\hline
0&[1,0,0,0,\dots,0,1]&(1,1,1,\dots,1,0)&2&$n-1$&1&$h_a{}^b$\\
1&[0,0,1,0,\dots,0,1]&(0,0,0,\dots,0,1)&2&1&1&$\tilde{A}_{(n-3,1)}$\\
1&[0,1,0,0,\dots,0,0]&(0,0,1,\dots,1,1)&0&$n-2$&0&$\tilde{A}_{(n-2)}$
\end{tabular}
\caption{\label{apppdec}The first level of $A_{n-3}^{+++}$
decomposed with respect to its regular $A_{n-1}$ subalgebra obtained by
deleting node $n$ which gives rise to a grading by $l$.}
\end{table}

We have included the dual of a (Cartan subalgebra) scalar
which is an allowed representation but
with vanishing outer multiplicity as we know from section
\ref{dectechnique}. Hence, very-extended $A^{+++}_{n-3}$ only contains
the graviton field and its dual. This supports the claim of \cite{We03a}
that this algebra lies behind a non-linear realisation of pure gravity
in $n$ dimensions. The content is in agreement with the oxidised
theory which is pure Einstein-Hilbert theory \cite{CrJuLuePo99}.
We propose a general result for level $2$ in the appendix \ref{apppl2}
which is also a non-trivial result about the general structure of this
family of very-extended Kac-Moody algebras.

\end{subsection}

\end{section}
%
%

\begin{section}{Non simply-laced cases}

We next turn to the study of very-extended algebras that derive from
classical algebras which are not simply-laced, starting with the
exceptional cases. Afterwards, the families of very-extended algebras
belonging to the $B$ and the $C$ series will be considered.

\begin{subsection}{$F_4^{+++}$}

The Dynkin diagram of $F_4^{+++}$ is depicted in
figure \ref{f4pppdynk}. The maximal gravity subalgebra is
five-dimensional and so the fields will be representations of
$\mathfrak{sl}(6)$.  The
list of representations occuring up to level $(2,4)$ (which is the
level of the affine root in affinised $F_4$) is in table \ref{f4pppdec}.
The lengths of the roots are normalised such that long roots have norm
square 2.

\begin{figure}
\begin{center}
\includegraphics{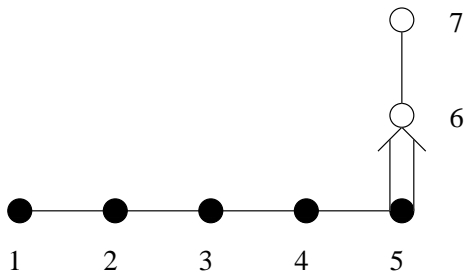}
\caption{\label{f4pppdynk}The Dynkin diagram of $F_4^{+++}$ drawn
with its $A_5$ subalgebra as gravity given by the solid nodes.}
\end{center}
\end{figure}

\begin{table}
\begin{tabular}{cccrrrr}
$(l_1,l_2)$&$A_5$ weight&$F_4^{+++}$ element
$\alpha$&$\alpha^2$&$ht(\alpha)$&$\mu$&Interpretation\\
\hline
(0,0)&[1,0,0,0,1]&(1,1,1,1,1,0,0)&2&5&1&$h_a{}^b$\\
(0,0)&[0,0,0,0,0]&(0,0,0,0,0,0,0)&0&0&1&$\phi$\\
(1,0)&[0,0,0,0,0]&(0,0,0,0,0,0,1)&1&1&1&$\chi$\\
(0,1)&[0,0,0,0,1]&(0,0,0,0,0,1,0)&1&1&1&$A_{(1)}$\\
(1,1)&[0,0,0,0,1]&(0,0,0,0,0,1,1)&1&2&1&$A_{(1)}$\\
(0,2)&[0,0,0,1,0]&(0,0,0,0,1,2,0)&2&3&1&$A_{(2)}$\\
(1,2)&[0,0,0,1,0]&(0,0,0,0,1,2,1)&1&4&1&$A_{(2)}$\\
(2,2)&[0,0,0,1,0]&(0,0,0,0,1,2,2)&2&5&1&$A_{(2)}$\\
(1,3)&[0,0,1,0,0]&(0,0,0,1,2,3,1)&1&7&1&$\tilde{A}_{(3)}$\\
(2,3)&[0,0,1,0,0]&(0,0,0,1,2,3,2)&1&8&1&$\tilde{A}_{(3)}$\\
(2,4)&[0,0,1,0,1]&(0,0,0,1,2,4,2)&2&9&1&$\tilde{A}_{(3,1)}$\\
(1,4)&[0,1,0,0,0]&(0,0,1,2,3,4,1)&1&11&1&$\tilde{A}_{(4)}^\chi$\\
(2,4)&[0,1,0,0,0]&(0,0,1,2,3,4,2)&0&12&1&$\tilde{A}_{(4)}^\phi$
\end{tabular}
\caption{\label{f4pppdec}The first levels of $F_4^{+++}$
decomposed with respect to its regular $A_5$ subalgebra obtained by
deleting nodes $7$ and $6$ giving a grading by $l_1$ and $l_2$
respectively.}
\end{table}

Examining the table we find that the fields up to and including the
level of the affine
root consist of gravity $h_a{}^b$ and a scalar $\phi$ on level
$(0,0)$, where $\phi$ lives in the Cartan subalgebra. On the positive
levels we find two one-form fields, and
one two-form field all together with  their duals as well
as one further two-form whose field strength is
self-dual. This self-duality follows from  the
listed fields as we have three two-form fields of which at most two
may be paired together to form a gauge field whose   field strength
has no self-duality condition. This content is the field content of
the
$F_4^{+++}$ oxidised theory listed in \cite{CrJuLuePo99}.

It is useful to precisely identify this theory. The bosonic content of the
(1,0) supersymmetry multiplets in six dimensions are as follows; the
supergravity mulitplet which has a graviton and  second rank tensor gauge
field, the vector multiplet which  has just a vector gauge field and
finally the tensor multiplet which has a scalar and a second rank tensor
gauge field. Both of the  second rank tensor gauge fields have field
strengths that are  self-dual with the one in
 the gravity multiplet being of opposite self-duality
to that in  the tensor multiplet. As such,  it is clear that the
$F_4^{+++}$ theory consist of supergravity coupled to two vector
multiplets and two tensor multiplets. The coupling of these multiplets is
given in  \cite{RiSa98}.

\end{subsection}

\begin{subsection}{$G_2^{+++}$}

The Dynkin diagram of very extended $G_2$ is shown in figure
\ref{g2pppdynk} and we list the representations in table
\ref{g2pppdec} up to height $6$.

\begin{figure}
\begin{center}
\includegraphics{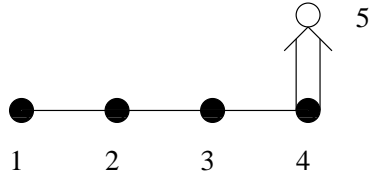}
\caption{\label{g2pppdynk}The Dynkin diagram of $G_2^{+++}$ drawn
with its $A_4$ subalgebra as gravity given by the solid nodes.}
\end{center}
\end{figure}

\begin{table}
\begin{tabular}{cccrrrr}
$l$&$A_4$ weight&$G_2^{+++}$ element
$\alpha$&$\alpha^2$&$ht(\alpha)$&$\mu$&Interpretation\\
\hline
0&[1,0,0,1]&(1,1,1,1,0)&2&4&1&$h_a{}^b$\\
1&[0,0,0,1]&(0,0,0,0,1)&2&1&1&$A_{(1)}$\\
2&[0,0,1,0]&(0,0,0,1,2)&2&3&1&$\tilde{A}_{(2)}$\\
3&[0,0,1,1]&(0,0,0,1,3)&6&4&1&$\tilde{A}_{(2,1)}$\\
3&[0,1,0,0]&(0,0,1,2,3)&0&6&0&$\tilde{A}_{(3)}^\phi$
\end{tabular}
\caption{\label{g2pppdec}The first levels of $G_2^{+++}$
decomposed with respect to its regular $A_4$ subalgebra obtained by
deleting nodes $5$ corresponding to a grading by $l$.}
\end{table}

All other representations occur at higher level. We see
that we have a five-dimensional
theory with a gravitational field and its dual and also
a 1-form and its dual. This is the right bosonic field content for
Einstein-Maxwell theory or $N=2$ supergravity in five dimensions
\cite{CrJuLuePo99}.
We list higher levels in appendix \ref{g2pppl2}.

\end{subsection}

\begin{subsection}{$B$ series}
\label{bppp}

The Dynkin diagram of $B_{n-3}^{+++}$ is
depicted in figure \ref{bpppdynk}. Similar to the $D$ series we face
the problem of expanding about two nodes but we can again apply our
low-level techniques to get the representations up to the level of the
affine root of the $B$ series. We denote the level as indicated in the
Dynkin diagram by $(l_1,l_2)$ corresponding to the entries on the
nodes $n$ and $n-1$ respectively. The affine root is the element
$(0,0,1,2,2,\dots,2,1)$ in $B_{n-3}^{+++}$ which is at level $(1,2)$
and so we have to construct the representations up to this level. By
similar arguments to section \ref{dppp} and taking antisymmetry into
account we find table \ref{bpppdec}.

\begin{figure}
\begin{center}
\includegraphics{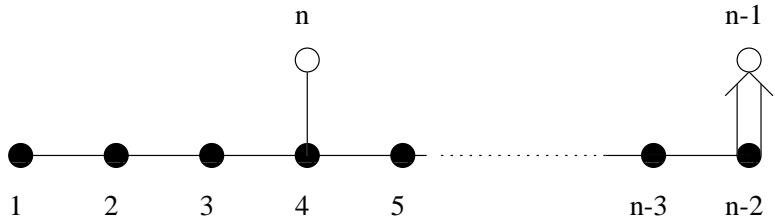}
\caption{\label{bpppdynk}The Dynkin diagram of $B_{n-3}^{+++}$ drawn
with its $A_{n-2}$ subalgebra as gravity given by the solid nodes.}
\end{center}
\end{figure}

\begin{table}
\begin{tabular}{cccrrrr}
$(l_1,l_2)$&$A_{n-2}$ weight&$B_{n-3}^{+++}$ element
$\alpha$&$\alpha^2$&$ht(\alpha)$&$\mu$&\\
\hline
(0,0)&[1,0,0,0,0,\dots,0,0,1]&(1,1,1,1,\dots,1,0,0)&2&$n-2$&1&$h_a{}^b$\\
(0,0)&[0,0,0,0,0,\dots,0,0,0]&(0,0,0,0,\dots,0,0,0)&0&0&1&$\phi$\\
(0,1)&[0,0,0,0,0,\dots,0,0,1]&(0,0,0,0,\dots,0,1,0)&1&1&1&$A_{(1)}$\\
(0,2)&[0,0,0,0,0,\dots,0,1,0]&(0,0,0,0,\dots,1,2,0)&2&3&1&$A_{(2)}$\\
(1,0)&[0,0,0,1,0,\dots,0,0,0]&(0,0,0,0,\dots,0,0,1)&2&1&1&$\tilde{A}_{(n-5)}$\\
(1,1)&[0,0,1,0,0,\dots,0,0,0]&(0,0,0,1,\dots,1,1,1)&1&$n-3$&1&$\tilde{A}_{(n-4)}$\\
(1,2)&[0,0,1,0,0,\dots,0,0,1]&(0,0,0,1,\dots,1,2,1)&2&$n-2$&1&$\tilde{A}_{(n-4,1)}$\\
(1,2)&[0,1,0,0,0,\dots,0,0,0]&(0,0,1,2,\dots,2,2,1)&0&$2n-6$&1&$\tilde{A}_{(n-3)}^\phi$
\end{tabular}
\caption{\label{bpppdec}The first levels of $B_{n-3}^{+++}$
decomposed with respect to its regular $A_{n-2}$ subalgebra obtained by
deleting nodes $n$ and $n-1$ giving a grading by $l_1$ and $l_2$
respectively.}
\end{table}

The field content up to the relevant level is thus a 1-form, a 2-form,
the metric and a dilaton and their duals, forming part of an
$(n-1)$-dimensional theory. This field content agrees with the oxidised
theory considered in  
\cite{CrJuLuePo99}. In ten dimensions it is just the bosonic field
content of $N=1$ supergravity plus one abelian vector multiplet.  The
non-linear realisation has been studied in \cite{SchnWe03}. Again,
there should be a different theory from the bifurcation
in the Dynkin diagram. The case of $B_3^{+++}$ proves to be
interesting and is analysed at higher levels in appendix
\ref{b3pppl2}.

\end{subsection}

\begin{subsection}{$C$ series}

The prescription for finding the field content of a tentative physical
theory applied to $C_{n-3}^{+++}$ tells us that we should start by
considering its $A_3$ subalgebra as indicated in figure \ref{cpppdynk}.

\begin{figure}
\begin{center}
\includegraphics{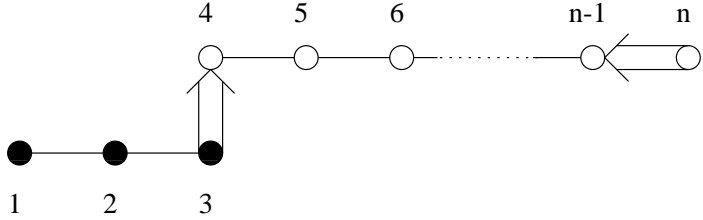}
\caption{\label{cpppdynk}The Dynkin diagram of $C_{n-3}^{+++}$ drawn
with its $A_3$ subalgebra as gravity given by the solid nodes.}
\end{center}
\end{figure}

We see that there is a $C_{n-3}$ subalgebra coupled via (the affine)
node $3$ to
this $A_3$ and so from the structure of $C_{n-3}$ this tells us that
there will be $(n-4)^2$ scalar fields and their duals
two-forms and $2n-8$ one-forms. We also find the metric and $n-4$ dilatons
and their dual fields up to the level of the affine $C$ root. This
agrees with the results in \cite{CrJuLuePo99}. We do
not list higher levels for any $C$ series case because of lack of
space and the large
number of fields already at low levels.

\end{subsection}

\end{section}

%
%

\begin{section}{Discussion}

\begin{subsection}{The $A_{d-3}^{+++}$ subalgebras}

The theories ${\cal V}_\lag$ based on the very-extended Kac-Moody
algebras $\lag^{+++}$
 all  contain a gravitational sector. It has been conjectured that pure
gravity in $d$-dimensional space-time  can be extended such that it is  a
non-linear realisation based on $A_{d-3}^{+++}$ i.e ${\cal V}_{A_{d-3}}$. 
To be consistent we would therefore have to find that $A_{d-3}^{+++}$
is contained in $\lag^{+++}$ for the appropriate $d$. This is known to
be true for $E_8^{+++}$ \cite{We03a}, but we now show it is true in all
cases. This can be understood by recalling Dynkin's
construction of all regular semi-simple subalgebras of $\lag$. We begin
considering   the Dynkin diagram of the non-twisted affine
$\lag^+$ and deleting nodes from it \cite{Dy57}. If we want to consider
${\cal V}_\lag$ as a
$d$-dimensional theory then we can find an $A_{d-1}$ regular
subalgebra in $\lag^{+++}$ which starts at the very-extended node. By
removing the over- and the very-extended node we obtain a $A_{d-3}$
subalgebra of $\lag^{+++}$. This is actually a subalgebra of $\lag^+$
whose Dynkin diagram is a subdiagram of the non-twisted affine one and
due to Dynkin's construction it is thus a subalgebra of $\lag$. Now we
just
have to very-extend both $\lag$ and this $A_{d-3}$ to find that there
exists $A_{d-3}^{+++}\subset \lag^{+++}$. Explicitly, we take as the
simple root extending $A_{d-1}$ to $A_{d-3}^{+++}$ the element
generating the dual graviton. It is encouraging that the non-linear
realisation of ${\cal V}_\lag$ contains the expected ${\cal
  V}_{A_{d-3}}$. 

We also note that table \ref{apppdec} tells us that if there are Cartan
subalgebra scalars $\phi$, their duals $A_{(d-2)}^\phi$
will come in at the same level as the
dual graviton. This has been observed in the case by case analysis too.

\end{subsection}

\begin{subsection}{T-duality and bifurcations}

In this section we propose a connection between the bifurcations (or
T-junctions) of the
Dynkin diagram and a generalisation of
T-duality transformations in the theory of gravity
and matter
associated with it. Suppose we are given a certain very-extended
algebra $\lag^{+++}$ with a choice of gravity subalgebra $A_{d-1}$. Then
Kaluza-Klein reductions can be realised by shortening the 
gravity subalgebra by a node from the far end
because this corresponds to decomposing
tensors of $\mathfrak{gl}(d)$ into tensors of $\mathfrak{gl}(d-1)$.
For example, this is how
the content of the {\rm IIA} decomposition of $E_8^{+++}$ can be
obtained from the $11d$ decomposition of $E_8^{+++}$, cf. figures
\ref{e8pppdynk} and \ref{e8pppabdynk}. If the Dynkin diagram of
$\lag^{+++}$ contains a bifurcation and our gravity subalgebra, called
$A_{d_A-1}$, stretches past it, it is possible to reduce the theory down
to the point of the bifurcation. The spectrum obtained then will
coincide with the spectrum of any other choice of gravity subalgebra
$A_{d_B-1}$ stretching past the bifurcation, subsequently reduced to
the bifurcation point.

For example, the $E_8^{+++}$ diagram has a bifurcation point and so we
can take different embeddings of gravity subalgebras and the theories
agree if reduced to nine dimensions. In particular, there are two
choices of $A_9$ subalgebra stretching past the bifurcation point and
the field contents of the different decompositions have been related
to {\rm IIA} and {\rm IIB} supergravity in section \ref{e8ppp}.
This agrees with the well-known feature of the two maximal ten-dimensional
supergravity theories that they are related by a T-duality
if each is compactified on a circle. In this case
the bifurcation
in the Dynkin diagrams gives rise to theories which are related by some
generalised T-duality transforms if dimensionally reduced down
to the bifurcation point. The proposal that $E_8^{+++}$ is a symmetry
not only of $11d$ supergravity but also of M-theory and thus of its
$10d$ superstring limit is consistent with this observation because it
is well-known that T-duality lifts to string theory.

In a similar way we can look at the $E_7^{+++}$. The bifurcation point
is the node labelled $6$ in figure \ref{e7pppdynk} and so we expect two
theories which agree when reduced to seven dimensions. 
One corresponds to the maximal ten-dimensional theory, called
interpretation (a) in section \ref{e7ppp}
while the other is the eight-dimensional interpretation (b). 
Naturally, both reduced theories 
are contained in the decomposition of $E_8^{+++}$ into
its $A_6$ subalgebra.

A similar proposal for generalising T-duality has
been put forward in \cite{Ke02}.

\end{subsection}

\begin{subsection}{Concluding remarks}

We have used the very-extended Kac-Moody
algebras $\lag^{+++}$ to introduce a set of theories ${\cal V}_\lag$ 
which are just the non-linear realisation based on $\lag^{+++}$. 
We might expect that the theories ${\cal
V}_\lag$ contain the oxidised theories
${\cal O}_\lag$ as subsectors since the latter by construction have a
symmetry group $\lag$ in three dimensions. In this paper we have
calculated the $A_{d-1}$ representations of $\lag^{+++}$ at low levels
and so deduced the field content of ${\cal V}_\lag$. It was found that
the fields of the theory ${\cal V}_\lag$ up to the level of the affine
root contain the fields of ${\cal O}_\lag$ in the correct
representation of $A_{d-1}$. This implies that the field content of
the theories ${\cal O}_\lag$ can be read off from the Dynkin diagram,
particularly simple at level $1$. We have also argued that the Dynkin
diagram encodes information about generalised T-dualities.

While it is an 
important consistency check that the field contents
match between ${\cal V}_\lag$ and ${\cal O}_\lag$ at low levels, it is
an important task to further uncover the structure of the full ${\cal
  V}_\lag$ which includes finding the correct dynamical equations
for the higher level fields, as listed in the appendices,
which are expected to be largely governed by the structure of
$\lag^{+++}$.

In this way of viewing things considered here, 
gravity can be embedded in a larger theory in a number of different
ways, some of which involve no supersymmetry. For example, in any
space-time  dimension,  $d$, gravity appears as part of the theories
${\cal V}_{A_{d-3}}$ and
${\cal V}_{D_{d-2}}$.  At low energy,  in the former theory one only has
gravity,  but in the latter one also has a dilaton and a
second rank tensor gauge field and in both of these theories one might
expect to see an infinite number of massive modes corresponding to the
infinite number of generators in the Kac-Moody algebra. In certain
other dimensions, for example,  in  ten and eleven dimensions one
has further possiblilties such as the theories associated with
${\cal V}_{E_{8}}$.

It is instructive to realise that all the theories that have been thought
to play a part in a unified theory of physics are included in this
picture based on very-extended Kac-Moody algebras and it is natural to
suppose that they are all part of some larger algebraic structure that
may be a  Borcherds algebra. This idea is consistent
with the old suggestion \cite{CaEnNiTa85,EnHoTa01}
that  the closed bosonic string might contain the superstrings. An
encouraging sign is that    $D_{24}^{+++}$
contains
$E_{8}^{+++}\times D_{16}$ if one includes the weight lattice of the
former
\cite{We02a}.\footnote{This remark is a slightly
corrected version of that given in this reference and we thank F.  Englert
for discussions on this point.}

The suggestion given above fits naturally into the historical
developments to find a unified theory. Although supergravity in
the late 1970s and early 1980s was hoped to provide a unified theory of
gravity and quantum theory it was  realised in the mid 1980s that
embedding these theories in,  the previously discovered,   superstrings
was more likely to be successful. However, by the  late 1990s it became
clear  that  superstring theories themselves were but part of some
bigger structure which was called M theory.  From the perspective
of the above suggestion,  M theory is itself just one step on the way and
that the very-extended algebras play an important r\^ole in fixing the
structure of that theory, similar to supersymmetry. 
The final algebraic structure may include Borcherds'
Fake
Monster algebra which is the vertex algebra on the unique even,
self-dual Lorentzian lattice in $26$ dimensions \cite{Bo90}. Its 
Dynkin diagram contains an
infinite number of nodes and it has been studied in the context of
string scattering amplitudes in \cite{Mo93}.

Finally we note that at present,
the theories ${\cal V}_\lag$ have been introduced as
non-linear realisations, very much along the same lines as elementary
particles where modelled in the early days before this was understood
as the breaking of a symmetry. It seems interesting to consider
what the unbroken theory looks like for ${\cal V}_\lag$.\\

\end{subsection}

\end{section}

{\bf Acknowledgements}\\
We are grateful for discussions with P. Goddard and M. R. Gaberdiel.
AK would like to thank King's College, London, for hospitality and the
EPSRC and the
Studienstiftung des deutschen Volkes for financial support. 
IS is grateful for the support by the Israel Academy of Science and
Humanities - Centers of Excellence Program, and the German-Israel
Bi-National Science Foundation. This
research was supported in part by the PPARC grants 
PPA/G/O/2000/00451  and PPA/G/S4/1998/00613. 

\newpage

\appendix

\begin{section}{Some results on higher levels}

In the appendices we list results obtained on the field contents of
the theories ${\cal V}_\lag$ if read in terms of the indicated $A$
subalgebras. Most of the results were obtained with the help of a
computer and give a flavour of the structure of $\lag^{+++}$ at higher
level. Lack of space prevents us from presenting more data.

\begin{subsection}{{\rm IIA} and {\rm IIB} theories from $E_8^{+++}$}
\label{e8pppl2}

The higher levels of the decomposition of $E_8^{+++}$ with respect to
its $A_{10}$ subalgebra are well-documented in the literature
\cite{NiFi03} to which we refer the reader. Here we give higher levels
of the two decompositions into $A_9$ subalgebras discussed in the
section \ref{e8ppp}.

{\bf IIA}: This is the data obtained for the {\rm IIA} case. All the
listed
levels are complete.

\begin{longtable}{cccrrr}
$(l_1,l_2)$&$A_9$ weight&$E_8^{+++}$ element
$\alpha$&$\alpha^2$&$ht(\alpha)$&$\mu$\\
\hline
\endhead
(1,0)&[0,0,0,0,0,0,0,1,0]&(0,0,0,0,0,0,0,0,0,0,1)&2&1&1\\
(0,1)&[0,0,0,0,0,0,0,0,1]&(0,0,0,0,0,0,0,0,0,1,0)&2&1&1\\
(1,1)&[0,0,0,0,0,0,1,0,0]&(0,0,0,0,0,0,0,1,1,1,1)&2&4&1\\
(2,1)&[0,0,0,0,1,0,0,0,0]&(0,0,0,0,0,1,2,3,2,1,2)&2&11&1\\
(3,1)&[0,0,1,0,0,0,0,0,0]&(0,0,0,1,2,3,4,5,3,1,3)&2&22&1\\
(4,1)&[1,0,0,0,0,0,0,0,0]&(0,1,2,3,4,5,6,7,4,1,4)&2&37&1\\
(2,2)&[0,0,0,1,0,0,0,0,0]&(0,0,0,0,1,2,3,4,3,2,2)&2&17&1\\
(3,2)&[0,0,1,0,0,0,0,0,1]&(0,0,0,1,2,3,4,5,3,2,3)&2&23&1\\
(3,2)&[0,1,0,0,0,0,0,0,0]&(0,0,1,2,3,4,5,6,4,2,3)&0&30&1\\
(4,2)&[0,1,0,0,0,0,0,1,0]&(0,0,1,2,3,4,5,6,4,2,4)&2&31&1\\
(4,2)&[1,0,0,0,0,0,0,0,1]&(0,1,2,3,4,5,6,7,4,2,4)&0&38&1\\
(4,2)&[0,0,0,0,0,0,0,0,0]&(1,2,3,4,5,6,7,8,5,2,4)&-2&47&2\\
(5,2)&[1,0,0,0,0,0,1,0,0]&(0,1,2,3,4,5,6,8,5,2,5)&2&41&1\\
(5,2)&[0,0,0,0,0,0,0,1,0]&(1,2,3,4,5,6,7,8,5,2,5)&0&48&1\\
(6,2)&[0,0,0,0,0,1,0,0,0]&(1,2,3,4,5,6,8,10,6,2,6)&2&53&1\\
(3,3)&[0,1,0,0,0,0,0,0,1]&(0,0,1,2,3,4,5,6,4,3,3)&2&31&1\\
(3,3)&[1,0,0,0,0,0,0,0,0]&(0,1,2,3,4,5,6,7,5,3,3)&0&39&0\\
(4,3)&[0,1,0,0,0,0,1,0,0]&(0,0,1,2,3,4,5,7,5,3,4)&2&34&1\\
(4,3)&[1,0,0,0,0,0,0,0,2]&(0,1,2,3,4,5,6,7,4,3,4)&2&39&1\\
(4,3)&[1,0,0,0,0,0,0,1,0]&(0,1,2,3,4,5,6,7,5,3,4)&0&40&1\\
(4,3)&[0,0,0,0,0,0,0,0,1]&(1,2,3,4,5,6,7,8,5,3,4)&-2&48&2\\
(5,3)&[0,1,0,0,1,0,0,0,0]&(0,0,1,2,3,5,7,9,6,3,5)&2&41&1\\
(5,3)&[1,0,0,0,0,0,1,0,1]&(0,1,2,3,4,5,6,8,5,3,5)&2&42&1\\
(5,3)&[1,0,0,0,0,1,0,0,0]&(0,1,2,3,4,5,7,9,6,3,5)&0&45&1\\
(5,3)&[0,0,0,0,0,0,0,1,1]&(1,2,3,4,5,6,7,8,5,3,5)&0&49&1\\
(5,3)&[0,0,0,0,0,0,1,0,0]&(1,2,3,4,5,6,7,9,6,3,5)&-2&51&3\\
(4,4)&[1,0,0,0,0,0,1,0,0]&(0,1,2,3,4,5,6,8,6,4,4)&2&43&1\\
(4,4)&[0,0,0,0,0,0,0,0,2]&(1,2,3,4,5,6,7,8,5,4,4)&2&49&1\\
(4,4)&[0,0,0,0,0,0,0,1,0]&(1,2,3,4,5,6,7,8,6,4,4)&0&50&0\\
(5,4)&[1,0,0,0,0,1,0,0,1]&(0,1,2,3,4,5,7,9,6,4,5)&2&46&1\\
(5,4)&[0,1,0,1,0,0,0,0,0]&(0,0,1,2,4,6,8,10,7,4,5)&2&47&1\\
(5,4)&[1,0,0,0,1,0,0,0,0]&(0,1,2,3,4,6,8,10,7,4,5)&0&50&1\\
(5,4)&[0,0,0,0,0,0,1,0,1]&(1,2,3,4,5,6,7,9,6,4,5)&0&52&2\\
(5,4)&[0,0,0,0,0,1,0,0,0]&(1,2,3,4,5,6,8,10,7,4,5)&-2&55&2\\
(5,5)&[1,0,0,1,0,0,0,0,0]&(0,1,2,3,5,7,9,11,8,5,5)&2&56&1\\
(5,5)&[0,0,0,0,0,1,0,0,1]&(1,2,3,4,5,6,8,10,7,5,5)&2&56&1\\
(5,5)&[0,0,0,0,1,0,0,0,0]&(1,2,3,4,5,7,9,11,8,5,5)&0&60&0\\
\end{longtable}

We note that there is a nine-form potential $A_{(9)}$ (supporting a
domain wall $D8$-brane in string theory) appearing at level
$(4,1)$ which probably pertains to massive {\rm IIA} supergravity
\cite{Ro86}. This field has no obvious dual in $10d$.
In \cite{SchnWe02} it was shown that massive {\rm IIA} supergravity
can be formulated as a non-linear realisation based on $E_8^{+++}$
with such a field where one does not need to include a dual.

{\bf IIB}: Below is the data obtained for the {\rm IIB} case.

\begin{longtable}{cccrrr}
$(l_1,l_2)$&$A_9$ weight&$E_8^{+++}$ element
$\alpha$&$\alpha^2$&$ht(\alpha)$&$\mu$\\
\hline
\endhead
(1,0)&[0,0,0,0,0,0,0,0,0]&(0,0,0,0,0,0,0,0,0,1,0)&2&1&1\\
(0,1)&[0,0,0,0,0,0,0,1,0]&(0,0,0,0,0,0,0,0,1,0,0)&2&1&1\\
(1,1)&[0,0,0,0,0,0,0,1,0]&(0,0,0,0,0,0,0,0,1,1,0)&2&2&1\\
(1,2)&[0,0,0,0,0,1,0,0,0]&(0,0,0,0,0,0,1,2,2,1,1)&2&7&1\\
(1,3)&[0,0,0,1,0,0,0,0,0]&(0,0,0,0,1,2,3,4,3,1,2)&2&16&1\\
(2,3)&[0,0,0,1,0,0,0,0,0]&(0,0,0,0,1,2,3,4,3,2,2)&2&17&1\\
(1,4)&[0,1,0,0,0,0,0,0,0]&(0,0,1,2,3,4,5,6,4,1,3)&2&29&1\\
(2,4)&[0,0,1,0,0,0,0,0,1]&(0,0,0,1,2,3,4,5,4,2,2)&2&23&1\\
(2,4)&[0,1,0,0,0,0,0,0,0]&(0,0,1,2,3,4,5,6,4,2,3)&0&30&1\\
(3,4)&[0,1,0,0,0,0,0,0,0]&(0,0,1,2,3,4,5,6,4,3,3)&2&31&1\\
(1,5)&[0,0,0,0,0,0,0,0,0]&(1,2,3,4,5,6,7,8,5,1,4)&2&46&1\\
(2,5)&[0,1,0,0,0,0,0,1,0]&(0,0,1,2,3,4,5,6,5,2,3)&2&31&1\\
(2,5)&[1,0,0,0,0,0,0,0,1]&(0,1,2,3,4,5,6,7,5,2,3)&0&38&1\\
(2,5)&[0,0,0,0,0,0,0,0,0]&(1,2,3,4,5,6,7,8,5,2,4)&-2&47&2\\
(3,5)&[0,1,0,0,0,0,0,1,0]&(0,0,1,2,3,4,5,6,5,3,3)&2&32&1\\
(3,5)&[1,0,0,0,0,0,0,0,1]&(0,1,2,3,4,5,6,7,5,3,3)&0&39&1\\
(3,5)&[0,0,0,0,0,0,0,0,0]&(1,2,3,4,5,6,7,8,5,3,4)&-2&48&2\\
(4,5)&[0,0,0,0,0,0,0,0,0]&(1,2,3,4,5,6,7,8,5,4,4)&2&49&1\\
(2,6)&[1,0,0,0,0,0,1,0,0]&(0,1,2,3,4,5,6,8,6,2,4)&2&41&1\\
(2,6)&[0,0,0,0,0,0,0,1,0]&(1,2,3,4,5,6,7,8,6,2,4)&0&48&1\\
(3,6)&[0,1,0,0,0,1,0,0,0]&(0,0,1,2,3,4,6,8,6,3,4)&2&37&1\\
(3,6)&[1,0,0,0,0,0,0,1,1]&(0,1,2,3,4,5,6,7,6,3,3)&2&40&1\\
(3,6)&[1,0,0,0,0,0,1,0,0]&(0,1,2,3,4,5,6,8,6,3,4)&0&42&1\\
(3,6)&[0,0,0,0,0,0,0,0,2]&(1,2,3,4,5,6,7,8,6,3,3)&0&48&0\\
(3,6)&[0,0,0,0,0,0,0,1,0]&(1,2,3,4,5,6,7,8,6,3,4)&-2&49&3\\
(4,6)&[1,0,0,0,0,0,1,0,0]&(0,1,2,3,4,5,6,8,6,4,4)&2&43&1\\
(4,6)&[0,0,0,0,0,0,0,1,0]&(1,2,3,4,5,6,7,8,6,4,4)&0&50&1\\
(2,7)&[0,0,0,0,0,1,0,0,0]&(1,2,3,4,5,6,8,10,7,2,5)&2&53&1\\
(3,7)&[1,0,0,0,0,1,0,0,1]&(0,1,2,3,4,5,7,9,7,3,4)&2&45&1\\
(3,7)&[0,1,0,1,0,0,0,0,0]&(0,0,1,2,4,6,8,10,7,3,5)&2&46&1\\
(3,7)&[1,0,0,0,1,0,0,0,0]&(0,1,2,3,4,6,8,10,7,3,5)&0&49&1\\
(3,7)&[0,0,0,0,0,0,0,2,0]&(1,2,3,4,5,6,7,8,7,3,4)&2&50&1\\
(3,7)&[0,0,0,0,0,0,1,0,1]&(1,2,3,4,5,6,7,9,7,3,4)&0&51&1\\
(3,7)&[0,0,0,0,0,1,0,0,0]&(1,2,3,4,5,6,8,10,7,3,5)&-2&54&3\\
(4,7)&[1,0,0,0,0,1,0,0,1]&(0,1,2,3,4,5,7,9,7,4,4)&2&46&1\\
(4,7)&[0,1,0,1,0,0,0,0,0]&(0,0,1,2,4,6,8,10,7,4,5)&2&47&1\\
(4,7)&[1,0,0,0,1,0,0,0,0]&(0,1,2,3,4,6,8,10,7,4,5)&0&50&1\\
(4,7)&[0,0,0,0,0,0,0,2,0]&(1,2,3,4,5,6,7,8,7,4,4)&2&51&1\\
(4,7)&[0,0,0,0,0,0,1,0,1]&(1,2,3,4,5,6,7,9,7,4,4)&0&52&1\\
(4,7)&[0,0,0,0,0,1,0,0,0]&(1,2,3,4,5,6,8,10,7,4,5)&-2&55&3\\
(5,7)&[0,0,0,0,0,1,0,0,0]&(1,2,3,4,5,6,8,10,7,5,5)&2&56&1\\
(4,8)&[0,1,1,0,0,0,0,0,1]&(0,0,1,3,5,7,9,11,8,4,5)&2&53&1\\
\end{longtable}

On level $(1,5)$ there is a trivial representation which can be read
as $A_{(10)}$ form (supporting the $D9$ brane).
 We observe that for both the {\rm IIA} and the {\rm IIB} decomposition
the only roots which can have vanishing outer
multiplicity seem to be the null roots. This had been noted for the
$A_{10}$ decomposition of $E_8^{+++}$ in \cite{NiFi03}.

\end{subsection}

\begin{subsection}{$E_7^{+++}$ in nine dimensions}
\label{e7pppl2}

Here we present more details for the decomposition of $E_7^{+++}$ with
respect to its regular $A_8$ subalgebra obtained by deleting the nodes
$10$ ($l_1$) and $9$ ($l_2$) in figure \ref{e7pppdynk}. We consider the
nine-dimensional viewpoint in order to facilitate a possible Lagrangian
interpretation of the representations that occur,
cf. \cite{CrJuLuePo99}. The notation is the
same as in section \ref{e7ppp}. Again
the null roots seem to be the only ones which allow vanishing
outer multiplicity.

\begin{longtable}{cccrrr}
$(l_1,l_2)$&$A_8$ weight&$E_7^{+++}$ element
$\alpha$&$\alpha^2$&$ht(\alpha)$&$\mu$\\
\hline
\endhead
(1,0)&[0,0,0,0,0,1,0,0]&(0,0,0,0,0,0,0,0,0,1)&2&1&1\\
(2,0)&[0,0,1,0,0,0,0,0]&(0,0,0,1,2,3,2,1,0,2)&2&11&1\\
(3,0)&[1,0,0,0,0,0,0,1]&(0,1,2,3,4,5,3,1,0,3)&2&22&1\\
(3,0)&[0,0,0,0,0,0,0,0]&(1,2,3,4,5,6,4,2,0,3)&0&30&0\\
(4,0)&[0,0,0,0,0,1,0,0]&(1,2,3,4,5,6,4,2,0,4)&2&31&1\\
(5,0)&[0,0,1,0,0,0,0,0]&(1,2,3,5,7,9,6,3,0,5)&2&41&1\\
(6,0)&[1,0,0,0,0,0,0,1]&(1,3,5,7,9,11,7,3,0,6)&2&52&1\\
(6,0)&[0,0,0,0,0,0,0,0]&(2,4,6,8,10,12,8,4,0,6)&0&60&0\\
(0,1)&[0,0,0,0,0,0,0,1]&(0,0,0,0,0,0,0,0,1,0)&2&1&1\\
(1,1)&[0,0,0,0,1,0,0,0]&(0,0,0,0,0,1,1,1,1,1)&2&5&1\\
(2,1)&[0,0,1,0,0,0,0,1]&(0,0,0,1,2,3,2,1,1,2)&2&12&1\\
(2,1)&[0,1,0,0,0,0,0,0]&(0,0,1,2,3,4,3,2,1,2)&0&18&1\\
(3,1)&[0,1,0,0,0,1,0,0]&(0,0,1,2,3,4,3,2,1,3)&2&19&1\\
(3,1)&[1,0,0,0,0,0,0,2]&(0,1,2,3,4,5,3,1,1,3)&2&23&1\\
(3,1)&[1,0,0,0,0,0,1,0]&(0,1,2,3,4,5,3,2,1,3)&0&24&1\\
(3,1)&[0,0,0,0,0,0,0,1]&(1,2,3,4,5,6,4,2,1,3)&-2&31&2\\
(4,1)&[1,0,0,0,1,0,0,1]&(0,1,2,3,4,6,4,2,1,4)&2&27&1\\
(4,1)&[0,1,1,0,0,0,0,0]&(0,0,1,3,5,7,5,3,1,4)&2&29&1\\
(4,1)&[1,0,0,1,0,0,0,0]&(0,1,2,3,5,7,5,3,1,4)&0&31&1\\
(4,1)&[0,0,0,0,0,1,0,1]&(1,2,3,4,5,6,4,2,1,4)&0&32&2\\
(4,1)&[0,0,0,0,1,0,0,0]&(1,2,3,4,5,7,5,3,1,4)&-2&35&2\\
(2,2)&[0,1,0,0,0,0,0,1]&(0,0,1,2,3,4,3,2,2,2)&2&19&1\\
(2,2)&[1,0,0,0,0,0,0,0]&(0,1,2,3,4,5,4,3,2,2)&0&26&0\\
(3,2)&[0,1,0,0,1,0,0,0]&(0,0,1,2,3,5,4,3,2,3)&2&23&1\\
(3,2)&[1,0,0,0,0,0,1,1]&(0,1,2,3,4,5,3,2,2,3)&2&25&1\\
(3,2)&[1,0,0,0,0,1,0,0]&(0,1,2,3,4,5,4,3,2,3)&0&27&1\\
(3,2)&[0,0,0,0,0,0,0,2]&(1,2,3,4,5,6,4,2,2,3)&0&32&1\\
(3,2)&[0,0,0,0,0,0,1,0]&(1,2,3,4,5,6,4,3,2,3)&-2&33&2\\
(4,2)&[1,0,0,0,1,0,1,0]&(0,1,2,3,4,6,4,3,2,4)&2&29&1\\
(4,2)&[0,1,1,0,0,0,0,1]&(0,0,1,3,5,7,5,3,2,4)&2&30&1\\
(4,2)&[1,0,0,1,0,0,0,1]&(0,1,2,3,5,7,5,3,2,4)&0&32&2\\
(4,2)&[0,0,0,0,0,1,0,2]&(1,2,3,4,5,6,4,2,2,4)&2&33&1\\
(4,2)&[0,0,0,0,0,1,1,0]&(1,2,3,4,5,6,4,3,2,4)&0&34&1\\
(4,2)&[0,2,0,0,0,0,0,0]&(0,0,2,4,6,8,6,4,2,4)&0&36&1\\
(4,2)&[0,0,0,0,1,0,0,1]&(1,2,3,4,5,7,5,3,2,4)&-2&36&4\\
(4,2)&[1,0,1,0,0,0,0,0]&(0,1,2,4,6,8,6,4,2,4)&-2&37&3\\
(4,2)&[0,0,0,1,0,0,0,0]&(1,2,3,4,6,8,6,4,2,4)&-4&40&3\\
(3,3)&[1,0,0,0,1,0,0,0]&(0,1,2,3,4,6,5,4,3,3)&2&31&1\\
(3,3)&[0,0,0,0,0,0,1,1]&(1,2,3,4,5,6,4,3,3,3)&2&34&1\\
(3,3)&[0,0,0,0,0,1,0,0]&(1,2,3,4,5,6,5,4,3,3)&0&36&0\\
(4,3)&[1,0,0,1,0,0,1,0]&(0,1,2,3,5,7,5,4,3,4)&2&34&1\\
(4,3)&[0,2,0,0,0,0,0,1]&(0,0,2,4,6,8,6,4,3,4)&2&37&1\\
(4,3)&[0,0,0,0,1,0,0,2]&(1,2,3,4,5,7,5,3,3,4)&2&37&1\\
(4,3)&[1,0,1,0,0,0,0,1]&(0,1,2,4,6,8,6,4,3,4)&0&38&2\\
(4,3)&[0,0,0,0,1,0,1,0]&(1,2,3,4,5,7,5,4,3,4)&0&38&2\\
(4,3)&[0,0,0,1,0,0,0,1]&(1,2,3,4,6,8,6,4,3,4)&-2&41&3\\
(4,3)&[1,1,0,0,0,0,0,0]&(0,1,3,5,7,9,7,5,3,4)&-2&44&2\\
(4,3)&[0,0,1,0,0,0,0,0]&(1,2,3,5,7,9,7,5,3,4)&-4&46&3

\end{longtable}

\end{subsection}

\begin{subsection}{$E_6^{+++}$ in eight dimensions}
\label{e6pppl2}

The following table contains the details of the decomposition of
$E_6^{+++}$ with respect to its regular $A_7$ subalgebra at the next
few levels. The notation is the same as in section \ref{e6ppp}. Again the
null roots seem to be the only ones which allow vanishing
outer multiplicity.

\begin{longtable}{cccrrr}
$(l_1,l_2)$&$A_7$ weight&$E_6^{+++}$ element
$\alpha$&$\alpha^2$&$ht(\alpha)$&$\mu$\\
\hline
\endhead
(1,0)&[0,0,0,0,0,0,0]&(0,0,0,0,0,0,0,0,1)&2&1&1\\
(0,1)&[0,0,0,0,1,0,0]&(0,0,0,0,0,0,0,1,0)&2&1&1\\
(1,1)&[0,0,0,0,1,0,0]&(0,0,0,0,0,0,0,1,1)&2&2&1\\
(0,2)&[0,1,0,0,0,0,0]&(0,0,1,2,3,2,1,2,0)&2&11&1\\
(1,2)&[0,0,1,0,0,0,1]&(0,0,0,1,2,1,0,2,1)&2&7&1\\
(1,2)&[0,1,0,0,0,0,0]&(0,0,1,2,3,2,1,2,1)&0&12&1\\
(2,2)&[0,1,0,0,0,0,0]&(0,0,1,2,3,2,1,2,2)&2&13&1\\
(0,3)&[0,0,0,0,0,0,1]&(1,2,3,4,5,3,1,3,0)&2&22&1\\
(1,3)&[0,1,0,0,1,0,0]&(0,0,1,2,3,2,1,3,1)&2&13&1\\
(1,3)&[1,0,0,0,0,0,2]&(0,1,2,3,4,2,0,3,1)&2&16&1\\
(1,3)&[1,0,0,0,0,1,0]&(0,1,2,3,4,2,1,3,1)&0&17&1\\
(1,3)&[0,0,0,0,0,0,1]&(1,2,3,4,5,3,1,3,1)&-2&23&2\\
(2,3)&[0,1,0,0,1,0,0]&(0,0,1,2,3,2,1,3,2)&2&14&1\\
(2,3)&[1,0,0,0,0,0,2]&(0,1,2,3,4,2,0,3,2)&2&17&1\\
(2,3)&[1,0,0,0,0,1,0]&(0,1,2,3,4,2,1,3,2)&0&18&1\\
(2,3)&[0,0,0,0,0,0,1]&(1,2,3,4,5,3,1,3,2)&-2&24&2\\
(3,3)&[0,0,0,0,0,0,1]&(1,2,3,4,5,3,1,3,3)&2&25&1\\
(1,4)&[1,0,0,1,0,0,1]&(0,1,2,3,5,3,1,4,1)&2&20&1\\
(1,4)&[0,2,0,0,0,0,0]&(0,0,2,4,6,4,2,4,1)&2&23&1\\
(1,4)&[1,0,1,0,0,0,0]&(0,1,2,4,6,4,2,4,1)&0&24&1\\
(1,4)&[0,0,0,0,1,0,1]&(1,2,3,4,5,3,1,4,1)&0&24&2\\
(1,4)&[0,0,0,1,0,0,0]&(1,2,3,4,6,4,2,4,1)&-2&27&2\\
(2,4)&[1,0,0,0,1,1,0]&(0,1,2,3,4,2,1,4,2)&2&19&1\\
(2,4)&[0,1,1,0,0,0,1]&(0,0,1,3,5,3,1,4,2)&2&19&1\\
(2,4)&[1,0,0,1,0,0,1]&(0,1,2,3,5,3,1,4,2)&0&21&2\\
(2,4)&[0,0,0,0,0,1,2]&(1,2,3,4,5,2,0,4,2)&2&23&1\\
(2,4)&[0,2,0,0,0,0,0]&(0,0,2,4,6,4,2,4,2)&0&24&1\\
(2,4)&[0,0,0,0,0,2,0]&(1,2,3,4,5,2,1,4,2)&0&24&0\\
(2,4)&[1,0,1,0,0,0,0]&(0,1,2,4,6,4,2,4,2)&-2&25&3\\
(2,4)&[0,0,0,0,1,0,1]&(1,2,3,4,5,3,1,4,2)&-2&25&4\\
(2,4)&[0,0,0,1,0,0,0]&(1,2,3,4,6,4,2,4,2)&-4&28&3\\
(3,4)&[1,0,0,1,0,0,1]&(0,1,2,3,5,3,1,4,3)&2&22&1\\
(3,4)&[0,2,0,0,0,0,0]&(0,0,2,4,6,4,2,4,3)&2&25&1\\
(3,4)&[1,0,1,0,0,0,0]&(0,1,2,4,6,4,2,4,3)&0&26&1\\
(3,4)&[0,0,0,0,1,0,1]&(1,2,3,4,5,3,1,4,3)&0&26&2\\
(3,4)&[0,0,0,1,0,0,0]&(1,2,3,4,6,4,2,4,3)&-2&29&2
\end{longtable}

\end{subsection}

\begin{subsection}{$D_8^{+++}$ at higher level}
\label{d8pppl2}

Here we list the representations occuring for $D_8^{+++}$ decomposed
with respect to its $A_9$ subalgebra at the next few levels.

\begin{longtable}{cccrrr}
$(l_1,l_2)$&$A_9$ weight&$D_8^{+++}$ element $\alpha$&$\alpha^2$
&$ht(\alpha)$&$\mu$\\
\hline
\endhead
(1,0)&[0,0,0,1,0,0,0,0,0]&(0,0,0,0,0,0,0,0,0,0,1)&2&1&1\\
(2,0)&[1,0,0,0,0,0,1,0,0]&(0,1,2,3,2,1,0,0,0,0,2)&2&11&1\\
(2,0)&[0,0,0,0,0,0,0,1,0]&(1,2,3,4,3,2,1,0,0,0,2)&0&18&0\\
(3,0)&[0,0,0,1,0,0,0,1,0]&(1,2,3,4,3,2,1,0,0,0,3)&2&19&1\\
(3,0)&[1,1,0,0,0,0,0,0,1]&(0,1,3,5,4,3,2,1,0,0,3)&2&22&1\\
(3,0)&[0,0,1,0,0,0,0,0,1]&(1,2,3,5,4,3,2,1,0,0,3)&0&24&0\\
(3,0)&[2,0,0,0,0,0,0,0,0]&(0,2,4,6,5,4,3,2,1,0,3)&0&30&0\\
(3,0)&[0,1,0,0,0,0,0,0,0]&(1,2,4,6,5,4,3,2,1,0,3)&-2&31&1\\
(4,0)&[0,1,0,0,1,0,0,0,1]&(1,2,4,6,4,3,2,1,0,0,4)&2&27&1\\
(4,0)&[1,0,0,0,0,0,1,1,0]&(1,3,5,7,5,3,1,0,0,0,4)&2&29&1\\
(4,0)&[2,0,0,1,0,0,0,0,0]&(0,2,4,6,5,4,3,2,1,0,4)&2&31&1\\
(4,0)&[1,0,0,0,0,1,0,0,1]&(1,3,5,7,5,3,2,1,0,0,4)&0&31&1\\
(4,0)&[0,1,0,1,0,0,0,0,0]&(1,2,4,6,5,4,3,2,1,0,4)&0&32&1\\
(4,0)&[1,0,0,0,1,0,0,0,0]&(1,3,5,7,5,4,3,2,1,0,4)&-2&35&1\\
(4,0)&[0,0,0,0,0,0,0,2,0]&(2,4,6,8,6,4,2,0,0,0,4)&0&36&0\\
(4,0)&[0,0,0,0,0,0,1,0,1]&(2,4,6,8,6,4,2,1,0,0,4)&-2&37&1\\
(4,0)&[0,0,0,0,0,1,0,0,0]&(2,4,6,8,6,4,3,2,1,0,4)&-4&40&1\\
(0,1)&[0,0,0,0,0,0,0,1,0]&(0,0,0,0,0,0,0,0,0,1,0)&2&1&1\\
(1,1)&[0,0,1,0,0,0,0,0,1]&(0,0,0,1,1,1,1,1,0,1,1)&2&7&1\\
(1,1)&[0,1,0,0,0,0,0,0,0]&(0,0,1,2,2,2,2,2,1,1,1)&0&14&1\\
(2,1)&[1,0,0,0,0,1,0,0,1]&(0,1,2,3,2,1,1,1,0,1,2)&2&14&1\\
(2,1)&[0,1,0,1,0,0,0,0,0]&(0,0,1,2,2,2,2,2,1,1,2)&2&15&1\\
(2,1)&[1,0,0,0,1,0,0,0,0]&(0,1,2,3,2,2,2,2,1,1,2)&0&18&1\\
(2,1)&[0,0,0,0,0,0,1,0,1]&(1,2,3,4,3,2,1,1,0,1,2)&0&20&1\\
(2,1)&[0,0,0,0,0,1,0,0,0]&(1,2,3,4,3,2,2,2,1,1,2)&-2&23&2\\
(1,2)&[0,1,0,0,0,0,0,1,0]&(0,0,1,2,2,2,2,2,1,2,1)&2&15&1\\
(1,2)&[1,0,0,0,0,0,0,0,1]&(0,1,2,3,3,3,3,3,1,2,1)&0&22&1\\
(1,2)&[0,0,0,0,0,0,0,0,0]&(1,2,3,4,4,4,4,4,2,2,1)&-2&31&1\\
(2,2)&[1,0,0,0,1,0,0,1,0]&(0,1,2,3,2,2,2,2,1,2,2)&2&19&1\\
(2,2)&[0,1,1,0,0,0,0,0,1]&(0,0,1,3,3,3,3,3,1,2,2)&2&21&1\\
(2,2)&[1,0,0,1,0,0,0,0,1]&(0,1,2,3,3,3,3,3,1,2,2)&0&23&2\\
(2,2)&[0,0,0,0,0,1,0,0,2]&(1,2,3,4,3,2,2,2,0,2,2)&2&23&1\\
(2,2)&[0,0,0,0,0,1,0,1,0]&(1,2,3,4,3,2,2,2,1,2,2)&0&24&1\\
(2,2)&[0,0,0,0,1,0,0,0,1]&(1,2,3,4,3,3,3,3,1,2,2)&-2&27&3\\
(2,2)&[0,2,0,0,0,0,0,0,0]&(0,0,2,4,4,4,4,4,2,2,2)&0&28&1\\
(2,2)&[1,0,1,0,0,0,0,0,0]&(0,1,2,4,4,4,4,4,2,2,2)&-2&29&3\\
(2,2)&[0,0,0,1,0,0,0,0,0]&(1,2,3,4,4,4,4,4,2,2,2)&-4&32&3\\
(1,3)&[1,0,0,0,0,0,1,0,0]&(0,1,2,3,3,3,3,4,2,3,1)&2&25&1\\
(1,3)&[0,0,0,0,0,0,0,1,0]&(1,2,3,4,4,4,4,4,2,3,1)&0&32&1
\end{longtable}

\end{subsection}

\begin{subsection}{$A$ series at level $2$}
\label{apppl2}

Here we present the result for the decomposition of $A_{n-3}^{+++}$
with respect to its regular $A_{n-1}$ subalgebra at level $l=2$ in the
notation of section \ref{appp}. We assume $n$ to sufficiently large
($n\ge 8$) for the expressions below to make sense, otherwise some of the
representations collapse in the obvious way.

\begin{longtable}{cccrrr}
$l$&$A_{n-1}$ weight&$A_{n-3}^{+++}$ element
$\alpha$&$\alpha^2$&$ht(\alpha)$&$\mu$\\
\hline
\endhead
1&[0,0,1,0,0,0,0,\dots,0,0,1]&(0,0,0,0,0,0,\dots,0,0,1)&2&1&1\\
&[0,1,0,0,0,0,0,\dots,0,0,0]&(0,0,1,1,1,1,\dots,1,1,1)&0&$n-2$&0\\
2&[1,0,0,0,1,0,0,\dots,0,1,0]&(0,1,2,1,0,0,\dots,0,1,2)&2&7&1\\
&[0,0,0,0,0,1,0,\dots,0,0,2]&(1,2,3,2,1,0,\dots,0,0,2)&2&11&1\\
&[0,0,0,0,0,1,0,\dots,0,1,0]&(1,2,3,2,1,0,\dots,0,1,2)&0&12&0\\
&[0,1,1,0,0,0,0,\dots,0,0,1]&(0,0,1,1,1,1,\dots,1,1,2)&2&$n-1$&1\\
&[1,0,0,1,0,0,0,\dots,0,0,1]&(0,1,2,1,1,1,\dots,1,1,2)&0&$n+1$&1\\
&[0,0,0,0,1,0,0,\dots,0,0,1]&(1,2,3,2,1,1,\dots,1,1,2)&-2&$n+5$&1\\
&[0,2,0,0,0,0,0,\dots,0,0,0]&(0,0,2,2,2,2,\dots,2,2,2)&0&$2n-4$&0\\
&[1,0,1,0,0,0,0,\dots,0,0,0]&(0,1,2,2,2,2,\dots,2,2,2)&-2&$2n-3$&1\\
&[0,0,0,1,0,0,0,\dots,0,0,0]&(1,2,3,2,2,2,\dots,2,2,2)&-4&$2n$&0
\end{longtable}

We have obtained this result by evaluating for a few cases
the exterior product of the
representation at level $1$ with itself and quotienting out the ideal
which we know explicitly in all cases. Then it is possible to
extrapolate the result to arbitrary cases. We have checked that the
above representations are allowed representations and that the
dimensions match in the way they should. We thus believe that this
result is true for all very extended $A$ series
algebras.\footnote{This statement could be proved for instance by
showing that the above weights are all allowed solutions. We have
checked that  then
the above combination is the only possible combination of their
dimensions to yield the right answer at level $2$. That these are all
possible solutions has been checked on a computer up to $n=27$.
Alternatively, one
could try to prove that the exterior product of the level one
representation $[0,0,1,0,0,\dots,0,1]$
with itself is given by the six representations above
plus the two representations $[0,1,0,1,0,0,\dots,0,0,2]$ and
$[0,0,2,0,0,\dots,0,1,0]$ contained in the ideal.} This is the
first non-trivial general result about
an infinite family of very-extended algebras to the best of our knowledge.
We also point out that the observation in \cite{NiFi03} that for
$E_8^{+++}$ only
representations with affine highest weight
can have outer multiplicity zero does not carry over to the
very-extended $A$ series.

\end{subsection}

\begin{subsection}{$F_4^{+++}$ at higher levels}
\label{f4pppl2}

Below are the $A_5$ representation in $F_5^{+++}$ at the next few levels.

\begin{longtable}{cccrrr}
$(l_1,l_2)$&$A_5$ weight&$F_4^{+++}$ element
$\alpha$&$\alpha^2$&$ht(\alpha)$&$\mu$\\
\hline
\endhead
(1,0)&[0,0,0,0,0]&(0,0,0,0,0,0,1)&1&1&1\\
(0,1)&[0,0,0,0,1]&(0,0,0,0,0,1,0)&1&1&1\\
(1,1)&[0,0,0,0,1]&(0,0,0,0,0,1,1)&1&2&1\\
(0,2)&[0,0,0,1,0]&(0,0,0,0,1,2,0)&2&3&1\\
(1,2)&[0,0,0,1,0]&(0,0,0,0,1,2,1)&1&4&1\\
(2,2)&[0,0,0,1,0]&(0,0,0,0,1,2,2)&2&5&1\\
(1,3)&[0,0,1,0,0]&(0,0,0,1,2,3,1)&1&7&1\\
(2,3)&[0,0,1,0,0]&(0,0,0,1,2,3,2)&1&8&1\\
(1,4)&[0,1,0,0,0]&(0,0,1,2,3,4,1)&1&11&1\\
(2,4)&[0,0,1,0,1]&(0,0,0,1,2,4,2)&2&9&1\\
(2,4)&[0,1,0,0,0]&(0,0,1,2,3,4,2)&0&12&1\\
(3,4)&[0,1,0,0,0]&(0,0,1,2,3,4,3)&1&13&1\\
(1,5)&[1,0,0,0,0]&(0,1,2,3,4,5,1)&1&16&1\\
(2,5)&[0,1,0,0,1]&(0,0,1,2,3,5,2)&1&13&1\\
(2,5)&[1,0,0,0,0]&(0,1,2,3,4,5,2)&-1&17&2\\
(3,5)&[0,1,0,0,1]&(0,0,1,2,3,5,3)&1&14&1\\
(3,5)&[1,0,0,0,0]&(0,1,2,3,4,5,3)&-1&18&2\\
(4,5)&[1,0,0,0,0]&(0,1,2,3,4,5,4)&1&19&1\\
(1,6)&[0,0,0,0,0]&(1,2,3,4,5,6,1)&1&22&1\\
(2,6)&[0,1,0,1,0]&(0,0,1,2,4,6,2)&2&15&1\\
(2,6)&[1,0,0,0,1]&(0,1,2,3,4,6,2)&0&18&1\\
(2,6)&[0,0,0,0,0]&(1,2,3,4,5,6,2)&-2&23&3\\
(3,6)&[0,1,0,1,0]&(0,0,1,2,4,6,3)&1&16&1\\
(3,6)&[1,0,0,0,1]&(0,1,2,3,4,6,3)&-1&19&3\\
(3,6)&[0,0,0,0,0]&(1,2,3,4,5,6,3)&-3&24&3\\
(4,6)&[0,1,0,1,0]&(0,0,1,2,4,6,4)&2&17&1\\
(4,6)&[1,0,0,0,1]&(0,1,2,3,4,6,4)&0&20&1\\
(4,6)&[0,0,0,0,0]&(1,2,3,4,5,6,4)&-2&25&3\\
(2,7)&[1,0,0,1,0]&(0,1,2,3,5,7,2)&1&20&1\\
(2,7)&[0,0,0,0,1]&(1,2,3,4,5,7,2)&-1&24&2\\
(3,7)&[0,1,1,0,0]&(0,0,1,3,5,7,3)&1&19&1\\
(3,7)&[1,0,0,0,2]&(0,1,2,3,4,7,3)&1&20&1\\
(3,7)&[1,0,0,1,0]&(0,1,2,3,5,7,3)&-1&21&3\\
(3,7)&[0,0,0,0,1]&(1,2,3,4,5,7,3)&-3&25&5\\
(4,7)&[0,1,1,0,0]&(0,0,1,3,5,7,4)&1&20&1\\
(4,7)&[1,0,0,0,2]&(0,1,2,3,4,7,4)&1&21&1\\
(4,7)&[1,0,0,1,0]&(0,1,2,3,5,7,4)&-1&22&3\\
(4,7)&[0,0,0,0,1]&(1,2,3,4,5,7,4)&-3&26&5
\end{longtable}

\end{subsection}

\begin{subsection}{$G_2^{+++}$ at higher levels}
\label{g2pppl2}

Below are the $A_4$ representations occuring in $G_2^{+++}$ at the
next few levels. We again note that there elements of norm squared
different
from $0$ with vanishing outer multiplicity.

\begin{longtable}{cccrrr}
$l$&$A_4$ weight&$G_2^{+++}$ element
$\alpha$&$\alpha^2$&$ht(\alpha)$&$\mu$\\
\hline
\endhead
1&[0,0,0,1]&(0,0,0,0,1)&2&1&1\\
2&[0,0,1,0]&(0,0,0,1,2)&2&3&1\\
3&[0,0,1,1]&(0,0,0,1,3)&6&4&1\\
3&[0,1,0,0]&(0,0,1,2,3)&0&6&0\\
4&[0,1,0,1]&(0,0,1,2,4)&2&7&1\\
4&[1,0,0,0]&(0,1,2,3,4)&-4&10&0\\
5&[0,1,1,0]&(0,0,1,3,5)&2&9&1\\
5&[1,0,0,1]&(0,1,2,3,5)&-4&11&1\\
5&[0,0,0,0]&(1,2,3,4,5)&-10&15&0\\
6&[0,1,1,1]&(0,0,1,3,6)&6&10&1\\
6&[1,0,0,2]&(0,1,2,3,6)&0&12&0\\
6&[0,2,0,0]&(0,0,2,4,6)&0&12&0\\
6&[1,0,1,0]&(0,1,2,4,6)&-6&13&2\\
6&[0,0,0,1]&(1,2,3,4,6)&-12&16&1\\
7&[0,2,0,1]&(0,0,2,4,7)&2&13&1\\
7&[1,0,1,1]&(0,1,2,4,7)&-4&14&2\\
7&[1,1,0,0]&(0,1,3,5,7)&-10&16&2\\
7&[0,0,0,2]&(1,2,3,4,7)&-10&17&1\\
7&[0,0,1,0]&(1,2,3,5,7)&-16&18&2\\
8&[1,0,1,2]&(0,1,2,4,8)&2&15&1\\
8&[0,2,1,0]&(0,0,2,5,8)&2&15&1\\
8&[1,0,2,0]&(0,1,2,5,8)&-4&16&1\\
8&[1,1,0,1]&(0,1,3,5,8)&-10&17&4\\
8&[0,0,0,3]&(1,2,3,4,8)&-4&18&0\\
8&[0,0,1,1]&(1,2,3,5,8)&-16&19&4\\
8&[2,0,0,0]&(0,2,4,6,8)&-16&20&1\\
8&[0,1,0,0]&(1,2,4,6,8)&-22&21&4\\
9&[0,2,1,1]&(0,0,2,5,9)&6&16&1\\
9&[1,0,2,1]&(0,1,2,5,9)&0&17&1\\
9&[1,1,0,2]&(0,1,3,5,9)&-6&18&3\\
9&[0,3,0,0]&(0,0,3,6,9)&0&18&0\\
9&[1,1,1,0]&(0,1,3,6,9)&-12&19&5\\
9&[0,0,1,2]&(1,2,3,5,9)&-12&20&3\\
9&[2,0,0,1]&(0,2,4,6,9)&-18&21&4\\
9&[0,0,2,0]&(1,2,3,6,9)&-18&21&4\\
9&[0,1,0,1]&(1,2,4,6,9)&-24&22&9\\
9&[1,0,0,0]&(1,3,5,7,9)&-30&25&5
\end{longtable}

\end{subsection}

\begin{subsection}{$B_3^{+++}$ at higher levels}
\label{b3pppl2}

First we list the representations occuring for the decomposition of
$B_3^{+++}$ with respect to its $A_4$ subalgebra as in section
\ref{bppp} at the next few levels. $B_3^{+++}$ is a particularly
interesting case as it admits a regular $A_5$ subalgebra as well and so
the
theory can be seen as living in six dimensions. This is related to an
observation in \cite{CrJuLuePo99}.

\begin{longtable}{cccrrr}
$(l_1,l_2)$&$A_4$ weight&$B_3^{+++}$ element
$\alpha$&$\alpha^2$&$ht(\alpha)$&$\mu$\\
\hline
\endhead
(1,0)&[0,0,0,1]&(0,0,0,0,0,1)&2&1&1\\
(0,1)&[0,0,0,1]&(0,0,0,0,1,0)&1&1&1\\
(1,1)&[0,0,1,0]&(0,0,0,1,1,1)&1&3&1\\
(0,2)&[0,0,1,0]&(0,0,0,1,2,0)&2&3&1\\
(1,2)&[0,0,1,1]&(0,0,0,1,2,1)&2&4&1\\
(1,2)&[0,1,0,0]&(0,0,1,2,2,1)&0&6&1\\
(2,2)&[0,1,0,1]&(0,0,1,2,2,2)&2&7&1\\
(2,2)&[1,0,0,0]&(0,1,2,3,2,2)&0&10&0\\
(1,3)&[0,1,0,1]&(0,0,1,2,3,1)&1&7&1\\
(1,3)&[1,0,0,0]&(0,1,2,3,3,1)&-1&10&1\\
(2,3)&[0,1,1,0]&(0,0,1,3,3,2)&1&9&1\\
(2,3)&[1,0,0,1]&(0,1,2,3,3,2)&-1&11&2\\
(2,3)&[0,0,0,0]&(1,2,3,4,3,2)&-3&15&1\\
(3,3)&[1,0,1,0]&(0,1,2,4,3,3)&1&13&1\\
(3,3)&[0,0,0,1]&(1,2,3,4,3,3)&-1&16&1\\
(1,4)&[0,1,1,0]&(0,0,1,3,4,1)&2&9&1\\
(1,4)&[1,0,0,1]&(0,1,2,3,4,1)&0&11&1\\
(1,4)&[0,0,0,0]&(1,2,3,4,4,1)&-2&15&1\\
(2,4)&[0,1,1,1]&(0,0,1,3,4,2)&2&10&1\\
(2,4)&[1,0,0,2]&(0,1,2,3,4,2)&0&12&1\\
(2,4)&[0,2,0,0]&(0,0,2,4,4,2)&0&12&1\\
(2,4)&[1,0,1,0]&(0,1,2,4,4,2)&-2&13&4\\
(2,4)&[0,0,0,1]&(1,2,3,4,4,2)&-4&16&3\\
(3,4)&[0,2,0,1]&(0,0,2,4,4,3)&2&13&1\\
(3,4)&[1,0,1,1]&(0,1,2,4,4,3)&0&14&2\\
(3,4)&[1,1,0,0]&(0,1,3,5,4,3)&-2&16&3\\
(3,4)&[0,0,0,2]&(1,2,3,4,4,3)&-2&17&2\\
(3,4)&[0,0,1,0]&(1,2,3,5,4,3)&-4&18&4\\
(4,4)&[1,1,0,1]&(0,1,3,5,4,4)&2&17&1\\
(4,4)&[0,0,1,1]&(1,2,3,5,4,4)&0&19&1\\
(4,4)&[2,0,0,0]&(0,2,4,6,4,4)&0&20&0\\
(4,4)&[0,1,0,0]&(1,2,4,6,4,4)&-2&21&2
\end{longtable}

Below are the results of the decomposition with respect to $A_5$ by
deleting node number $5$ in the relevant diagram.

\begin{longtable}{cccrrr}
$l$&$A_5$ weight&$B_3^{+++}$ element
$\alpha$&$\alpha^2$&$ht(\alpha)$&$\mu$\\
\hline
\endhead
1&[0,0,0,1,0]&(0,0,0,0,1,0)&1&1&1\\
2&[0,0,1,0,1]&(0,0,0,1,2,0)&2&3&1\\
2&[0,1,0,0,0]&(0,0,1,2,2,1)&0&6&0\\
\hline
3&[0,1,0,1,0]&(0,0,1,2,3,1)&1&7&1\\
3&[1,0,0,0,1]&(0,1,2,3,3,1)&-1&10&1\\
3&[0,0,0,0,0]&(1,2,3,4,3,2)&-3&15&0\\
4&[0,1,1,0,1]&(0,0,1,3,4,1)&2&9&1\\
4&[1,0,0,1,1]&(0,1,2,3,4,1)&0&11&1\\
4&[0,2,0,0,0]&(0,0,2,4,4,2)&0&12&0\\
4&[1,0,1,0,0]&(0,1,2,4,4,2)&-2&13&2\\
4&[0,0,0,0,2]&(1,2,3,4,4,1)&-2&15&1\\
4&[0,0,0,1,0]&(1,2,3,4,4,2)&-4&16&1\\
5&[1,0,1,0,2]&(0,1,2,4,5,1)&1&13&1\\
5&[0,2,0,1,0]&(0,0,2,4,5,2)&1&13&1\\
5&[1,0,1,1,0]&(0,1,2,4,5,2)&-1&14&2\\
5&[1,1,0,0,1]&(0,1,3,5,5,2)&-3&16&3\\
5&[0,0,0,1,2]&(1,2,3,4,5,1)&-1&16&1\\
5&[0,0,0,2,0]&(1,2,3,4,5,2)&-3&17&2\\
5&[0,0,1,0,1]&(1,2,3,5,5,2)&-5&18&4\\
5&[2,0,0,0,0]&(0,2,4,6,5,3)&-5&20&1\\
5&[0,1,0,0,0]&(1,2,4,6,5,3)&-7&21&3\\
6&[1,0,1,2,0]&(0,1,2,4,6,2)&2&15&1\\
6&[0,2,1,0,1]&(0,0,2,5,6,2)&2&15&1\\
6&[1,1,0,0,3]&(0,1,3,5,6,1)&2&16&1\\
6&[1,0,2,0,1]&(0,1,2,5,6,2)&0&16&1\\
6&[1,1,0,1,1]&(0,1,3,5,6,2)&-2&17&4\\
6&[0,3,0,0,0]&(0,0,3,6,6,3)&0&18&0\\
6&[0,0,1,0,3]&(1,2,3,5,6,1)&0&18&1\\
6&[0,0,0,3,0]&(1,2,3,4,6,2)&0&18&0\\
6&[1,1,1,0,0]&(0,1,3,6,6,3)&-4&19&5\\
6&[0,0,1,1,1]&(1,2,3,5,6,2)&-4&19&6\\
6&[2,0,0,0,2]&(0,2,4,6,6,2)&-4&20&2\\
6&[2,0,0,1,0]&(0,2,4,6,6,3)&-6&21&5\\
6&[0,1,0,0,2]&(1,2,4,6,6,2)&-6&21&7\\
6&[0,0,2,0,0]&(1,2,3,6,6,3)&-6&21&4\\
6&[0,1,0,1,0]&(1,2,4,6,6,3)&-8&22&10\\
6&[1,0,0,0,1]&(1,3,5,7,6,3)&-10&25&9\\
6&[0,0,0,0,0]&(2,4,6,8,6,4)&-12&30&1\\
\end{longtable}

We see that this is theory in six dimensions contains a 2-form with
self-dual field strength and the metric and its dual which is just
$d=6$, $N=1$ supergravity, in agreement with \cite{CrJuLuePo99}.

\end{subsection}

\end{section}

\end{document}